\documentclass{aa}

\usepackage{graphics,graphicx}
\usepackage{xcolor}
\usepackage{amsmath}
\usepackage{amssymb}
\usepackage{braket}
\usepackage{wrapfig}
\usepackage{caption}
\usepackage{subcaption}
\usepackage{natbib}
\usepackage{longtable}
\usepackage{stfloats} 
\usepackage{soul} 
\usepackage{ulem}
\usepackage{threeparttable}
\usepackage{textcomp}
\usepackage{xurl}
\usepackage{xcolor}

\definecolor{CiteColor}{HTML}{3B6EA5} 
\definecolor{UrlColor}{HTML}{2F5D8A}  
\definecolor{LinkColor}{HTML}{2F3A4A} 

\usepackage[unicode=true,breaklinks=true,colorlinks=true]{hyperref}
\hypersetup{
  linkcolor=LinkColor,   
  citecolor=CiteColor,   
  urlcolor=UrlColor      
}

\providecommand{\doi}[1]{} 
\renewcommand{\doi}[1]{\href{https://doi.org/#1}{doi:\,#1}}

\providecommand{\adsurl}[1]{}
\renewcommand{\adsurl}[1]{\href{#1}{ADS}}

\providecommand{\eprint}[1]{}
\renewcommand{\eprint}[1]{\href{https://arxiv.org/abs/#1}{arXiv:#1}}

\newcommand\orcid[1]{\protect\href{http://orcid.org/#1}{\includegraphics[height=12pt]{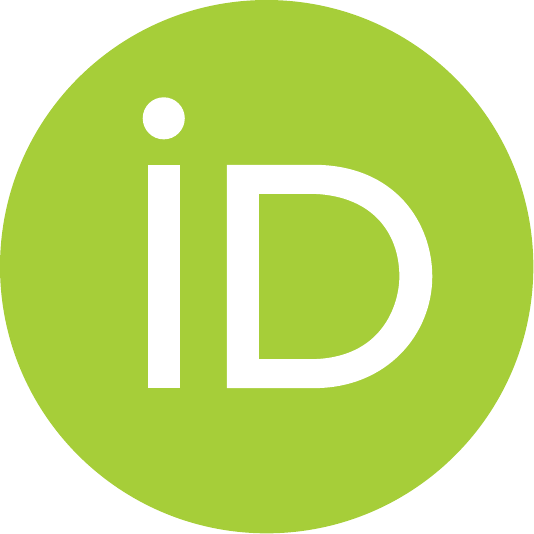}}}

\newcommand{\mstar}{M$_\star$}
\newcommand{\msun}{${\mathrm M}_\odot$}

\newcommand{\routflow}{$\dot{M}_{\rm outflow}$}
\newcommand{\rinfall}{$\dot{M}_{\rm infall}$}
\newcommand{\mgas}{${\rm M}_{\rm gas}$}
\newcommand{\f}{$f_\star$}
\newcommand{\Mup}{$m_{\rm up}$}

\newcommand{\mz}{$M_\star$--$Z$}

\definecolor{forestgreen}{rgb}{0.13, 0.55, 0.13}

\bibpunct{(}{)}{;}{a}{}{,} 

\begin{document}
\title{Limited imprint of high-mass IMF variations on sodium abundances in main-sequence galaxies}
\titlerunning{}

\author{Ziyi Guo\inst{1,2}\orcid{0000-0002-7532-1496}
\and Donatella Romano\inst{3}\orcid{0000-0002-0845-6171}
\and Zhiqiang Yan\inst{1,2}\orcid{0000-0001-7395-1198}
\and Zhi-Yu Zhang\inst{1,2}    \thanks{Corresponding author: \email{zzhang@nju.edu.cn}}       \orcid{0000-0002-7299-2876}
\and Xiaoting Fu\inst{4}\orcid{0000-0002-6506-1985}
\and Lizhi Xie\inst{5}\orcid{0000-0003-3864-068X}
\and Tereza Jerabkova\inst{6,8}\orcid{0000-0002-1251-9905}
\and Gabriella De Lucia\inst{9}\orcid{0000-0002-6220-9104}
\and Michaela Hirschmann\inst{10}\orcid{0000-0002-3301-3321}
\and Fabio Fontanot\inst{9,11}\orcid{0000-0003-4744-0188}
\and Eda Gjergo\inst{1,2}\orcid{0000-0002-7440-1080}
\and Alice Concas\inst{6,7}\orcid{0000-0003-3203-9818}
\and Xiaodong Tang\inst{12,13}\orcid{0000-0002-0029-4999}
}
\institute{School of Astronomy and Space Science, Nanjing University, Nanjing 210023, China\\
\email{zzhang@nju.edu.cn,zyguo@smail.nju.edu.cn}
\and Key Laboratory of Modern Astronomy and Astrophysics (Nanjing University), Ministry of Education, Nanjing 210023, China
\and INAF, Osservatorio di Astrofisica e Scienza dello Spazio, via Gobetti 93/3, 40129 Bologna, Italy
\and Purple Mountain Observatory, Chinese Academy of Sciences, Nanjing, China
\and Tianjin Normal University, Binshuixidao 393, 300387, Tianjin, China
\and European Southern Observatory, Karl-Schwarzschild-Strasse 2, 85748 Garching bei München, Germany
\and Scuola Normale Superiore, Piazza dei Cavalieri 7, 50126 Pisa, Italy
\and Department of Theoretical Physics and Astrophysics, Faculty of Science, Masaryk University, Kotláˇrská 2, Brno 611 37, Czech Republic
\and INAF-Astronomical Observatory of Trieste, via Riccardo Bozzoni, 2, I-34124 Trieste, Italy
\and DARK, Niels Bohr Institute, University of Copenhagen, Lyngbyvej 2, DK-2100 Copenhagen, Denmark
\and IFPU - Institute for Fundamental Physics of the Universe, via Beirut 2, I-34151, Trieste, Italy
\and Institute of Modern Physics, Chinese Academy of Sciences, Lanzhou 730000, China
\and School of Nuclear Science and Technology, University of Chinese Academy of Sciences, Beijing 100049, China
}
\date{}

\abstract{Growing evidence suggests that the stellar initial mass function (IMF)
varies systematically across galaxies, deviating from the canonical Milky Way
form. Such variations would modify the integrated nucleosynthetic yields, and
hence the abundance patterns used in stellar population synthesis studies. How
these could impact, in particular, the sodium abundance (and sodium-to-oxygen
ratios) in star-forming galaxies is not well understood. In this work, we
systematically study how high-mass IMF variations affect sodium enrichment using
a one-zone galactic chemical evolution model. The model incorporates star
formation histories from semi-analytic simulations and is calibrated to match
the observed galaxy mass--metallicity relation. We find that varying the IMF
high-mass end (and the IMF slope) could only alter the sodium abundance by less
than 0.1 dex, across galaxies with stellar masses from $10^9\,\mathrm{M}_\odot$
to $10^{11}\,\mathrm{M}_\odot$. This result is robust under different stellar
models and galaxy evolution assumptions, primarily because sodium production is
similar to that of oxygen. We conclude that sodium abundance is largely
insensitive to changes in the high-mass IMF, unlikely to compromise the use of
sodium indices as IMF diagnostics in stellar population studies.} 


\maketitle

\section{Introduction}
\label{sec:introduction}

The stellar initial mass function (IMF) describes the distribution of stellar
masses at birth in a single stellar population \citep[and references
therein]{1955ApJ...121..161S,jerabkova2025}. It is a basic assumption in many
fields of astrophysics, entering the derivation of the star formation rate
\citep[SFR,][]{kennicutt1998},  the mass-to-light ratios
\citep{bernardi2018,2024A&A...689A.221H}, and the star formation history
\citep[SFH,][]{madau2014,annibali2022} of galaxies. 

The IMF has long been treated as universal, motivated by observations in the
Milky Way and the Magellanic Clouds \citep{kroupa2001,kroupa2002,Bastian2010}.
This assumption greatly simplified the interpretation of galaxy properties.
However, accumulating observational and theoretical evidence now indicates that
the IMF varies systematically with galaxy properties and the physical
environments of star formation \citep{2018A&A...620A..39J,2026enap....2..173K}. 

In environments with high star-formation surface density, the IMF appears to
become more top-heavy (i.e. relatively biased to massive stars), as inferred
using independent diagnostics including H$\alpha$ equivalent-width--colour
relations \citep{2011MNRAS.415.1647G}, chemical abundance ratios \citep[e.g.
{$[\mathrm{Mg/Fe}]$}, and other
elements,][]{Ballero2007,2019A&A...632A.110Y,2021A&A...655A..19Y,jerabkova2025},
and molecular gas tracers such as CO isotopologue ratios \citep{henkel1993,
romano2017, zhang2018, guo2024}. Conversely, in low star-formation density
systems, the IMF is found to be more top-light, implying a relative deficit of
massive stars
\citep[e.g.][]{lee2009,2017A&A...607A.126Y,minelli2021,2025MNRAS.538.2989C}.

These trends are consistent with the integrated galactic initial mass function
theory, where the high-mass slope of the IMF depends on the galactic SFR
or SFR surface density
\citep{2003ApJ...598.1076K,2017A&A...607A.126Y,2018MNRAS.479.5678F}. A top-heavy
IMF in high-density environments is preferred from a theoretical perspective, in
which cosmic rays play a dominant role in heating the dense star-forming gas and
raise the Jeans mass \citep{2011MNRAS.414.1705P}. At low metallicity, magnetic
fields likely play an important role: the virial parameter systematically falls
below unity, suggesting that massive star formation becomes easier once
sufficient cloud mass is available \citep{Lin2025}. Such effects should be more
prominent at high redshift, reinforced by a higher cosmic microwave background
temperature, which sets a fundamental lower limit on the temperature of
star-formation gas \citep{Zhang2016,2022MNRAS.514.4639C}.

In contrast to the top-heavy IMF found in energetic environments, metal-rich
systems appear to favour the formation of low-mass stars. Star-counting analyses
within the Milky Way have suggested that the IMF varies with metallicity
\citep{kroupa2002,li2023}, with more low-mass stars towards higher
metallicities. Extending beyond the local neighbourhood, massive early-type
galaxies --- characterised by high metallicity --- exhibit spectral features
indicative of an enhanced fraction of low-mass stars compared to the canonical
IMF. This `bottom-heavy' IMF scenario is supported by numerous studies of galaxy
integrated spectra, dynamic studies and mass-to-light ratios
\citep[e.g.,][]{bruzual2003, conroy2012, labarbera2013, 2015ApJ...806L..31M,
vazdekis2016, maraston2020,Gu2022, Yan2024}.

Sodium, whose only stable isotope is \(^{23}\)Na, has the potential to be a
sensitive probe of the IMF, because its yields receive contributions mostly from
the massive stars, while also some contributions from the intermediate-mass
stars. In intermediate-mass stars, the dominant production channel is the Ne-Na
chain via $^{22}\text{Ne}(p, \gamma)^{23}\text{Na}$, mostly during the
asymptotic giant branch (AGB) phase. The required $^{22}\text{Ne}$ seed is from
the initial gas material that formed the star or was produced during the
preceding helium-burning stage through the chain $^{14}\text{N}(\alpha,
\gamma)^{18}\text{F}(e^+\nu)^{18}\text{O}(\alpha, \gamma)^{22}\text{Ne}$
\citep[see e.g. the nuclear reaction collection for stellar evolution in
][]{Fu2018}, which is heavily dependent on the N abundance. For significant
$^{23}$Na production, mixing processes, such as hot bottom burning
\citep[e.g.][]{Ventura2005, karakas2010} or rotation-induced mixing, are
necessary to bring protons into contact with this $^{22}\text{Ne}$ seed at
sufficiently high temperatures. In the absence of such mixing, the
$^{22}\text{Ne}$ in the helium-burning zones is instead primarily converted to
$^{25}\text{Mg}$ via the $^{22}\text{Ne}(\alpha, n)^{25}\text{Mg}$ reaction
during later evolutionary stages. In massive stars, besides $^{22}\text{Ne}(p,
\gamma)^{23}\text{Na}$, the hydrostatic Carbon burning becomes the primary
source of $^{23}$Na. The reaction $^{12}\text{C}(^{12}\text{C},
p)^{23}\text{Na}$ is one of the main fusion channels during this phase. Unlike
the material in the stellar core, the $^{23}$Na synthesised in the
Carbon-burning shells is preserved and efficiently ejected into the interstellar
medium via stellar winds or the subsequent supernova explosion. 

Observationally, several Na transitions can be routinely detected at
near-infrared $J$- and $K$-bands, which offer the advantage of being less
affected by dust extinction \citep{eftekhari2021}. Negative radial gradients of
the Na{\sc i} 2.21 $\mu$m line are frequently observed in elliptical galaxies
and galactic bulges \citep{alton2018, labarbera2025}. In the optical band,
\citet{concas2019} shows a positive trend between absorption strength of Na{\sc
i} D and both the star formation rate and the stellar mass, along with
signatures of contamination from interstellar absorption from neutral gas
outflows, introducing an additional source of uncertainty in the stellar
population analyses. A further complication arises from spectral blending:
\citet{rock2017} demonstrates that the Na{\sc i} 2.21 \textmu m index is contaminated
by weak molecular absorption lines, particularly from CN. Consequently, an
enhancement in $[\mathrm{C/Fe}]$ strengthens these molecular blends,
artificially inflating the measured Na index.

To shed light on the origin of Na and disentangle any dependence of its
abundance on the IMF, we investigate how different IMF assumptions and
nucleosynthesis prescriptions affect the evolution of the Na abundance using a
set of galactic chemical evolution (GCE) models. GCE acts as a bridge between
stellar nucleosynthesis and galaxy evolution by self-consistently accounting for
star formation, gas flows, and stellar feedback over cosmic
time\citep[e.g.,][]{Matteucci2012, Romano2010}. This approach enables a detailed
yet computationally efficient exploration of IMF-dependent abundance patterns
across a wide range of galaxy masses.

This paper is organised as follows. We describe the adopted GCE model in
Sect.~\ref{sec:model assumptions} and benchmark it against Milky Way
observations in Sect.~\ref{sec:model benchmark}. In Sect.~\ref{sec:models of
SF galaxies} we extend the model to star-forming galaxies with
$M_\star=10^9$--$10^{11}\,\mathrm{M}_\odot$. We discuss the implications in
Sect.~\ref{sec:discussion}. Throughout this paper, we adopt the Solar abundance
from \citep{Asplund2009} for abundance normalisation. 

\section{Model assumptions} \label{sec:model assumptions}

We use the one-zone chemical evolution code One-zone Model for the Evolution of
GAlaxies (OMEGA), distributed within the open-source NuGrid Python Chemical
Evolution Environment (NuPyCEE)
package \citep{cote2017}.  We
adopt the \texttt{SF mode} in this code, which allows us to control the gas mass
available to form stars at each time step through the adoption of a specific SFH
for each galaxy. The gas mass evolves as:

\begin{equation}
\label{equation: total gas evolution}
\begin{split}
     M_{\rm gas}&(t+\Delta t) = M_{\rm gas}(t) \\
                &\quad + [\dot{M}_{\rm infall}(t)-\dot{M}_{\rm outflow}(t)-\dot{M}_\star(t)+\dot{M}_{\rm ej}(t)] \cdot \Delta t,
\end{split}
\end{equation}

where $\Delta t$ is the time step.  At each time step, the gas mass is regulated
by four physical processes: gas accretion (\rinfall), gas loss (\routflow), star
formation ($\dot{M}_\star$), and mass return from dying stars ($\dot{M}_{\rm
ej}$). We assume instantaneous mixing within the single zone, and that gas loss
carry the ISM composition at the time of ejection.

In \texttt{SF mode}, the SFH is an input. For the Milky Way benchmark, we adopt
a simplified SFH profile inspired by the two-infall model of
\citet{chiappini1997}, without imposing a gas-density threshold for star
formation activity. For external star-forming galaxies, similarly robust SFH
constraints are not available;  we adopt SFHs extracted from the the GAlaxy
Evolution and Assembly (GAEA) semi-analytic model of hierarchical galaxy
formation \citep[see
Sect.~\ref{subsec:sfh}]{delucia2024,fontanot2017,fontanot2018,Fontanot2024}.

The gas reservoir mass at each timestep is inferred by inverting the (adapted)
Kennicutt-Schmidt star formation relation \citep{schmidt1959,kennicutt1998}
implemented in OMEGA, i.e. a linear relation between SFR and gas mass with given
star formation efficiency and timescale:

\begin{equation} \label{equation: k-s law}
        M_{\rm gas}(t) = \tau_\star \frac{\dot{M}_\star(t)}{\epsilon_\star},
\end{equation}

where $\epsilon_\star$ is the dimensionless star formation efficiency  and
$\tau_\star$ is the star formation timescale. These two quantities are assumed
to be constants in time and merged into a single parameter,

\begin{equation}
        f_\star = \frac{\epsilon_\star}{\tau_\star},
\end{equation}

which is a free time-invariant parameter that relates the gas mass to the SFR at
each time step. The present-day galaxy gas mass and SFR jointly constrain the
parameter \f.

The gas loss rate is related to the SFR via a mass-loading factor \(\eta\):

\begin{equation} \label{equation: outflow rate}
        \dot{M}_{\rm outflow}(t) = \eta \dot{M}_\star(t),
\end{equation}

where $\eta$ is assumed time-independent and is calibrated in
Sect.~\ref{sec:model benchmark} using the gas-phase oxygen abundance
constraints.

The infall rate is then determined by the adopted SFH, $\eta$, \f, and
$\dot{M}_{\rm ej}$:

\begin{equation}
\begin{split}
    \dot{M}_{\rm infall}(t) =& \frac{1}{f_\star\,\Delta t} [\dot{M}_\star(t+\Delta t)-\dot{M}_\star(t)] \\
    & + (\eta+1)\dot{M}_\star(t)-\dot{M}_{\rm ej}(t).
\end{split}
\end{equation} 

The gas mass returned by dying stars at each time step is computed by accounting
for finite stellar lifetimes. The ejection rates therefore depend on the SFH and
on the metallicity at the time of star formation. 

\subsection{Initial mass function}
\label{subsec:imf}

We adopt a three-part power-law IMF \citep{kroupa2001},

\begin{equation}
    \label{equation: IMF}
        \xi(m) = \frac{dN}{dm} = \left\{
    \begin{array}{ll}
            k_1 \times m^{-\alpha_1}, & m_{\rm low} \leq m < m_{\rm  1},\\
            k_2 \times m^{-\alpha_2}, & m_{\rm 1  } \leq m < m_{\rm  2},\\ 
            k_3 \times m^{-\alpha_3}, & m_{\rm 2  } \leq m < m_{\rm up},
    \end{array}
    \right.
\end{equation}

where \(m\) is in units of \msun.  The constants \(k_1\), \(k_2\), and \(k_3\)
are chosen to ensure continuity of \(\xi(m)\) at \(m_1\) and \(m_2\). \(m_{\rm
low}\) and \(m_{\rm up}\) are the lower and upper IMF boundaries of stellar
masses, and \(m_1\) and \(m_2\) are the IMF slope's break points.  

We adopt \(m_{\rm low}=0.01\,\mathrm{M}_\odot\), \(m_1=0.08\,\mathrm{M}_\odot\),
and \(m_2=0.5\,\mathrm{M}_\odot\), and vary the upper-mass limit \(m_{\rm up}\)
to assess the impact of the high-mass end of the IMF on the Na production. We
normalise the IMF such that \(\int_{m_{\rm low}}^{m_{\rm up}} m\,\xi(m)\,dm =
1\,\mathrm{M}_\odot\). The default slopes are
\((\alpha_1,\alpha_2,\alpha_3)=(0.3,1.3,2.3)\) \citep{kroupa2001}.

\subsection{Stellar yields}
\label{subsec:stellar yield}

\begin{figure}
    \centering
    \includegraphics[width=\columnwidth]{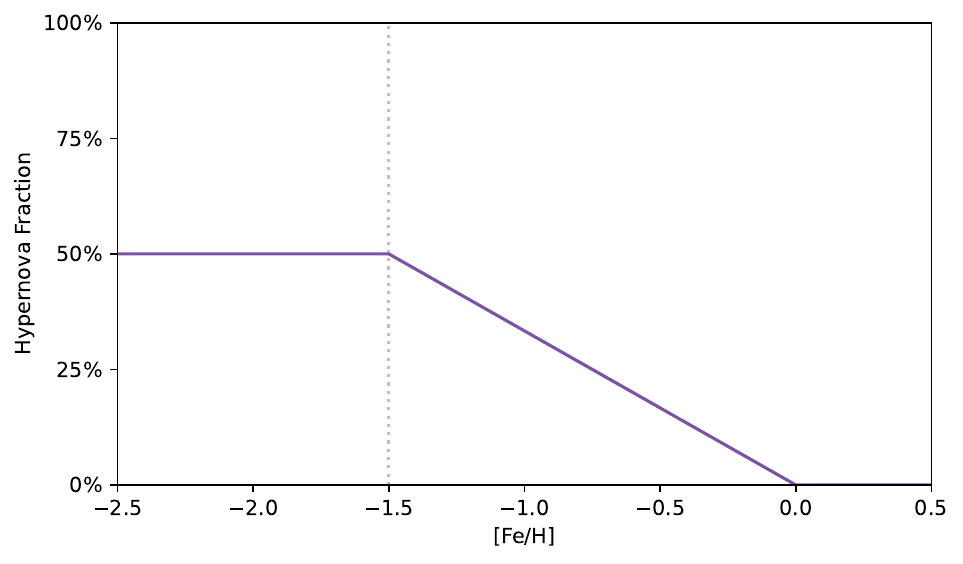}
        \caption{Hypernova fraction as a function of metallicity traced by
        $[\mathrm{Fe/H}]$, adopted in galactic chemical evolution (GCE) models by
        implementing N13 yields \citep[][]{nomoto2013} for massive stars,
        following the physical framework established by \citet{Kobayashi2006}.
        The fraction is set to \(f_{\rm HN}= 50\% \) for \([\mathrm{Fe/H}]<-1.5
        \), \(f_{\rm HN}=0 \) for \([\mathrm{Fe/H}] = 0 \), and varies linearly
        in between. }
    \label{fig: hypernova_frac}
\end{figure}

For low- and intermediate-mass stars (LIMS, initial masses
\(1\text{--}8\,\mathrm{M}_\odot\)), we adopt the yields of AGB stars from
\citet{karakas2010} and \citet{Cinquegrana2022}.  We combine the two yield sets
across mass and metallicity, ensuring a continuous coverage over the adopted AGB
mass range.  The nuclear reactions that synthesize and destroy $^{23}$Na are key
updates in \citet{karakas2010}, since the treatment of the Ne--Na cycle
reactions govern Na production and destruction in AGB envelopes. 

Type~Ia supernovae (SNe; singular SN) are implemented via a power-law delay-time
distribution \citep[DTD,][]{maoz2012}, which captures the delayed Fe enrichment
on Gyr timescales.  We adopt the Type~Ia SN yields of \citet[model
W7]{iwamoto1999}. The DTD normalisation and related assumptions are described in
Sect.~\ref{sec:model benchmark}. 

For massive stars, we consider three yield sets to assess the sensitivity of our
results to the choice of massive-star yields, since most sodium is produced by
massive stars. The baseline yield set for massive stars is taken from
\citet[][hereafter N13]{nomoto2013}. The N13 tables provide yields for massive
stars exploding either as normal core-collapse SNe (explosion energies
$\sim$10$^{51}$~ergs), or as hypernovae (with much higher explosion energies).
Following the physical framework established by \citet{Kobayashi2006}, where
hypernova contribution is dominant at low metallicities, we parameterize the
fraction of massive stars that explode as hypernovae as a linear function of
metallicity traced by $[\mathrm{Fe/H}]$ (Fig.~\ref{fig: hypernova_frac}): 

\begin{equation}
\label{equation:hn_fraction}
f_{\rm HN}([\mathrm{Fe/H}])=
  \begin{cases}
  0.5, & [\mathrm{Fe/H}]<-1.5,\\
  -\dfrac{1}{3}[\mathrm{Fe/H}], & -1.5 \le [\mathrm{Fe/H}] < 0,\\
  0, & [\mathrm{Fe/H}] \ge 0,
  \end{cases}
\end{equation}

The second and third massive-star yield sets are taken from \citet[][hereafter
LC18]{limongi2018}. We adopt their non-rotating models (\(v_{\rm
rot}=0\,\mathrm{km\,s^{-1}}\)) and consider two explosion prescriptions: the
recommended \texttt{set~R} and \texttt{set~M}. Both sets employ a mixing-and-
fallback scheme \citep{umeda2002}: the inner boundary of the mixed region is set
to reproduce \([\mathrm{Ni/Fe}] = 0.2\), the outer boundary is placed at the
base of the oxygen-burning shell, and the mass cutoff is chosen to eject
\(0.07\,\mathrm{M}_\odot\) of \(^{56}\)Ni.  The key difference between the two
models is that \texttt{set~M} assumes successful core-collapse supernova
explosions across the full massive-star mass range, whereas in \texttt{set~R}
all stars with \(m>25\,\mathrm{M}_\odot\) are assumed to collapse directly to
the stellar remnant. Consequently, the yields of the most massive stars in
\texttt{set~R} are only contributed by pre-SN stellar winds.

\begin{figure*}
    \centering
    \includegraphics[width=\linewidth]{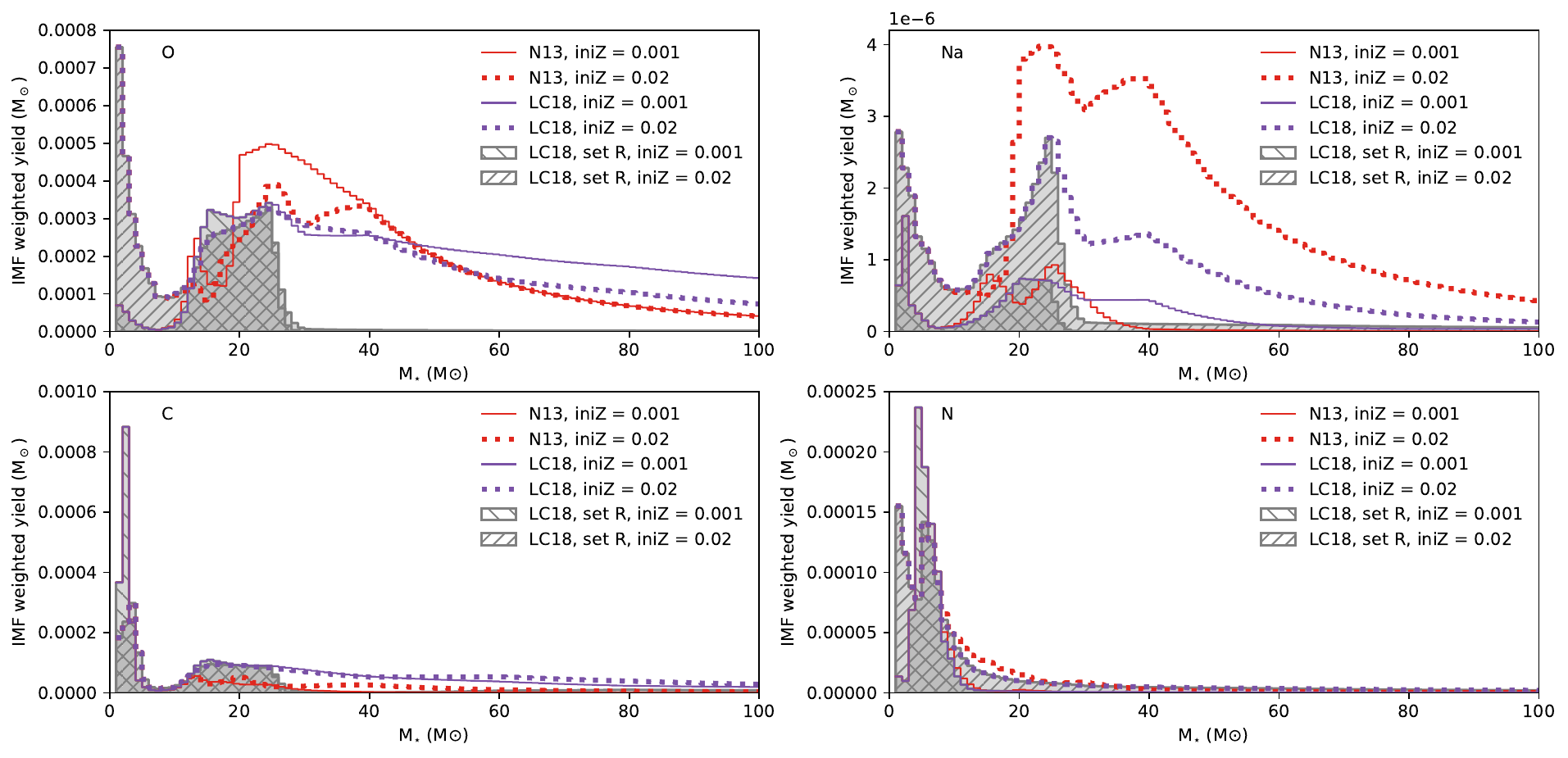}
    \caption{IMF-weighted yields (per unit initial stellar mass) of oxygen
        (upper-left), sodium (upper-right), carbon (lower-left), and nitrogen
        (lower-right). AGB yields are from \citet{karakas2010}. Massive-star
        yields are from \citet[][red lines]{nomoto2013} and \citet[][$v_{\rm
        rot}=0\,\mathrm{km\,s^{-1}}$; \texttt{set~M}: purple lines;
        \texttt{set~R}: grey histograms]{limongi2018}. Solid lines show models
        with initial metallicity \(Z=0.001\), and dotted lines show \(Z=0.02\).}
    \label{fig: yield}
\end{figure*}

The IMF-weighted yields of oxygen and sodium for the three yield sets adopted in
this study are shown in the top panels of Fig.~\ref{fig: yield}. The
contributions from different initial-mass bins are computed assuming the
canonical Kroupa IMF \citep{kroupa2001} and are normalised to an initial stellar
mass of \(1\,\mathrm{M}_\odot\). For N13 and LC18 \texttt{set~M} (\(v_{\rm
rot}=0\); red and purple curves, respectively), stars with
\(m>20\,\mathrm{M}_\odot\) produce more than \(\sim 70\%\) of the oxygen, with
only a weak dependence on metallicity (Fig.~\ref{fig: yield}, upper-left). AGB
stars contribute at most \(\sim 10\%\) of the total oxygen budget at \(Z\simeq
Z_\odot\).

The IMF- weighted sodium yield, on the other hand, increases with the initial
metallicity of the stellar population (Fig.~\ref{fig: yield}, upper-right).  As
an odd-Z element \citep{1971ApJ...166..153A,2006ApJ...653.1145K}, more than
\(\sim 80\%\) of sodium is produced by stars with \(m>20\,\rm{M}_\odot\) at
\(Z\simeq Z_\odot\), whereas at low metallicity this fraction decreases to
\(\sim 50\%\).

The grey histograms illustrate the effect of adopting LC18 \texttt{set~R}, in
which all stars with \(m>25\,\mathrm{M}_\odot\) are assumed to collapse directly
to the remnant. Compared to that of \texttt{set~M},  the oxygen production is
strongly suppressed, causing the predicted mass--metallicity relation to fall
below the observed one (see Sect.~\ref{sec:models of SF galaxies}). 

Previous Milky Way GCE studies that adopt \texttt{set~R} similarly underpredict
oxygen abundances \citep{2018MNRAS.476.3432P}. This discrepancy may arise
because the failed-SN or direct-collapse prescription is based on single-star
evolution, whereas interactions in massive binaries are common and may alter the
fraction of failed explosions and the amount of ejecta
\citep{2014ApJ...782....7D,2021ApJ...920....5L}. This could change the expected
fraction of failed SN and also promote the formation of an accretion disk and
outflow from the metal-rich fallback shell of a failed SN
\citep{2021ApJ...920....5L}.  For these reasons, in the following analysis we
only compare results obtained from N13 and  LC18 \texttt{set~M} $v =
0\,\mathrm{km\,s^{-1}}$.

Carbon and nitrogen are key intermediate products that set CNO abundances of
stellar cores and therefore the production of \(^{22}\)Ne, as a seed to produce
\(^{23}\)Na through \(\rm 22Ne(p,\gamma )23Na\)
\citep[e.g.,][]{Fu2018,Ventura2005}. We therefore also show the IMF-weighted
yields of C and N in the bottom panels of Fig.~\ref{fig: yield}. In our adopted
yield sets, the IMF-weighted carbon yield from AGB stars shows a strong
metallicity dependence, with higher C production at lower \(Z\). The nitrogen
yield exhibits a comparatively weak metallicity dependence over the two
metallicities shown in Fig.~\ref{fig: yield}.

\section{Model benchmark} \label{sec:model benchmark}

We first benchmark our model against Milky Way constraints. As described in
Sect.~\ref{sec:model assumptions}, the free parameters in the \texttt{SF mode}
are the star-formation-law parameter \(f_\star\) and the mass-loading factor
\(\eta\). We model the Solar neighbourhood as an annulus of width 2~kpc and
vertical thickness 1~kpc, centred at the present-day Solar radius. The final
adopted value of \(f_\star\) is $1.32\times10^{-10}$ $\mathrm{M}_\odot^{-1}$ and
the \(\eta\) is 1.0. We note that the “gas loss” in our model is not equivalent
to the broadly defined galactic outflow. In the Milky Way-calibrated setup, we
model only the annulus of the solar neighbourhood centred at the present-day
solar radius. As a result, our definition of gas loss also includes gas that is
removed via radial flows, which can lead to a systematically different inferred
mass-loading factor compared to measurements defined for the entire galaxy.

We adopt an SFH profile simplified from that of \citet{Chiappini2001}. This
profile is obtained by converting their SFR surface-density evolution into an
absolute SFR, by multiplying by the Solar-neighbourhood area. We slightly
rescale this SFH to match the present-day stellar surface density to the updated
value of $25.25\,\rm{M}_\odot\cdot pc^{-2}$, which is inferred from Eq.~(13) of
\citet{hunter2024} (instead of the \citet{Gilmore1989} value adopted in
\citealt{Chiappini2001}). The present-day gas mass surface density is
$22.23\,\rm{M}_\odot \cdot pc^{-2}$, which is  obtained by integrating Eq.~(14)
in \citet{hunter2024}. Given the present-day gas mass and SFR, we derive
\(f_\star\) (equivalent to gas depletion time) for the adopted linear
star-formation law. Our reference IMF is the Kroupa IMF \citep{kroupa2001}. 

The Type~Ia SN rate is implemented via a power-law DTD, following \citet{maoz2012}:

\begin{equation} \label{equation: DTD}
 \Psi(t) = A \left(\frac{t}{\mathrm{Gyr}}\right)^{\beta},
\end{equation}

where \(\Psi(t)\) is the number of SNe~Ia per unit time and per unit formed
stellar mass (\(\mathrm{yr^{-1}\,M_\odot^{-1}}\)). The normalisation factor
\(A\) is set by requiring

\begin{equation} \label{equation: DTD_norm}
  \int_{t_{\min}}^{t_{\max}} \Psi(t)\,dt = \frac{N_{\rm Ia}}{M_\star},
\end{equation}

We adopt \(N_{\rm Ia}/M_\star = 2.5\times10^{-3}\,\mathrm{M_\odot^{-1}}\) and
\(\beta=-1.12\) \citep{marinacci2014}. The integral is computed from $t = 0$ to
$t = 13.8\,\rm{Gyr}$.

\begin{figure*}
  \centering
  \includegraphics[width=2\columnwidth]{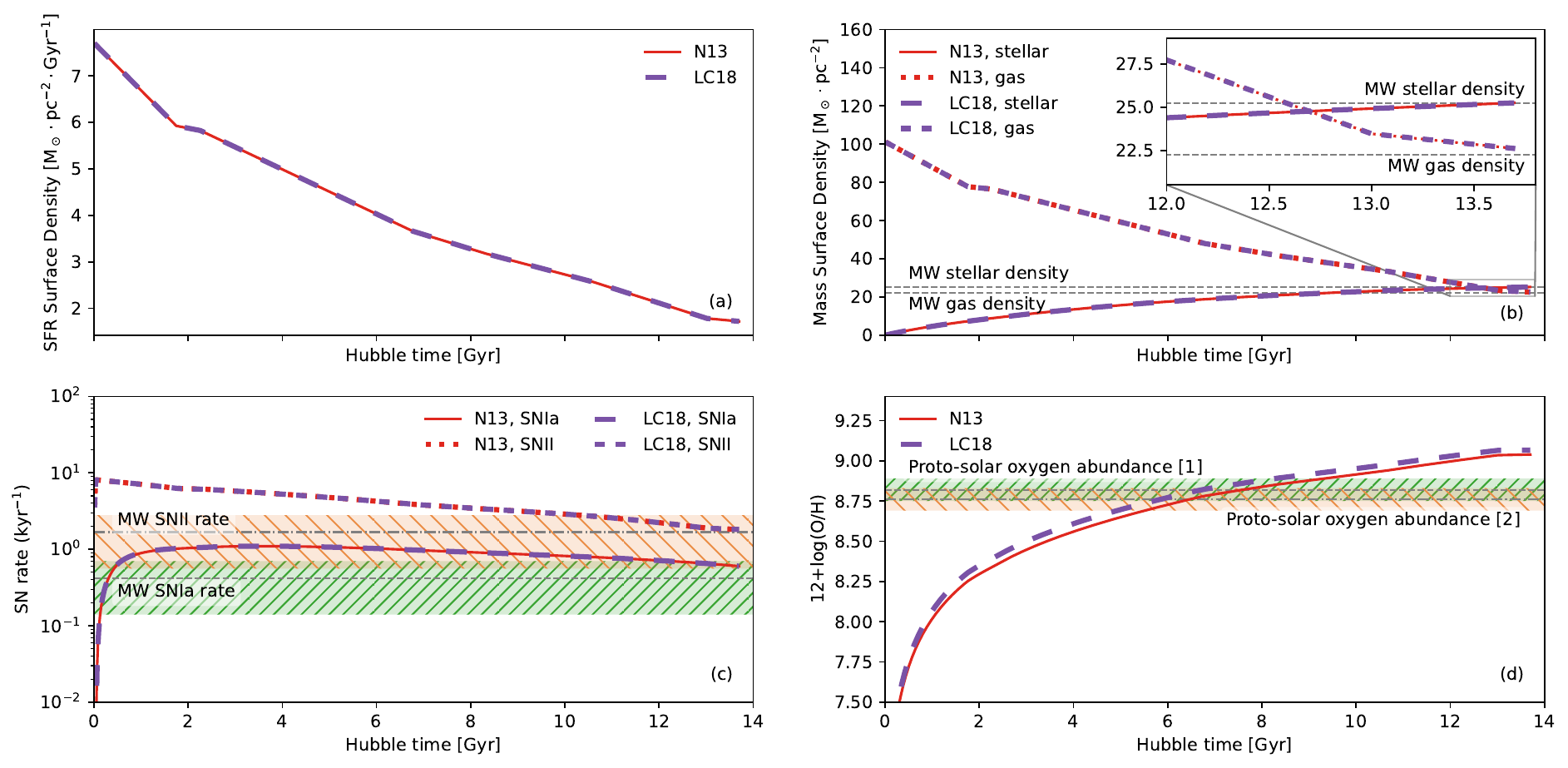} 
  \caption{Evolution of the SFR (upper-left), surface densities of living stars
        and gas (upper-right), SN rates (lower-left), and metallicity traced by
        gas-phase oxygen abundance (lower-right) for the Milky Way benchmark
        model. Results are shown for two massive-star yield sets: N13 (red
        lines) and LC18 \texttt{set~M} with \(v_{\rm
        rot}=0\,\mathrm{km\,s^{-1}}\) (purple lines).  Present-day stellar and
        gas surface densities are from \citet{hunter2024}; observed SN rates are
        from \citet{Cappellaro1999}; Solar oxygen abundances from
        \citet{Asplund2009,Asplund2021,Lodders2019,caffau2008} are shown for
        comparison.}\label{fig: MWmodel1}
\end{figure*}

Figure~\ref{fig: MWmodel1} shows the evolution of several key quantities for two
choices of massive-star yields, for the Milky Way benchmark model. In both
cases, AGB yields are from \citet{karakas2010}, while massive-star yields are
from N13 (red) or LC18 set~M with \(v_{\rm rot}=0\,\mathrm{km\,s^{-1}}\)
(purple). The SFR and the surface densities of living stars (Stars that are
alive at the present time, excluding stellar remnants) and gas are shown in the
upper panels, and the SN rates in the lower-left panel. These global quantities
are essentially insensitive to the adopted nucleosynthesis prescriptions,
whereas the oxygen abundance (lower-right panel of Fig.~\ref{fig: MWmodel1})
shows only a weak dependence on the choice of yields, with slightly higher oxygen 
abundance for the LC18 \texttt{set~M} yield.

The gas mass closely follows the imposed SFR because the gas
reservoir is inferred by inverting the adopted star-formation law (see
Sect.~\ref{sec:model assumptions}). The grey dashed lines in the upper-right
panel of Fig.~\ref{fig: MWmodel1} indicate the present-day stellar and gas
surface densities inferred by \citet{hunter2024}. Our model approaches these
values at the final stage of the evolution.

The model also reproduces the observed present-day Type~II and Type~Ia SN rates
within the uncertainty bar \citep[orange and green shaded regions in
Fig.~\ref{fig: MWmodel1}, lower-left panel;][]{Cappellaro1999}. The predicted
protosolar abundance of oxygen, 4.6~Gyr ago from the present day, is
\(12+\log(\mathrm{O/H})=8.82\pm0.07\), slightly higher than the values
(\(8.69\pm0.05\)) recommended by \citet{Asplund2009,Asplund2021} and
\citet{Lodders2019} , but consistent with the \(8.76\pm0.07\) estimate from
\citet{caffau2008}.

\begin{figure*}  
  \centering
  \includegraphics[width=2\columnwidth]{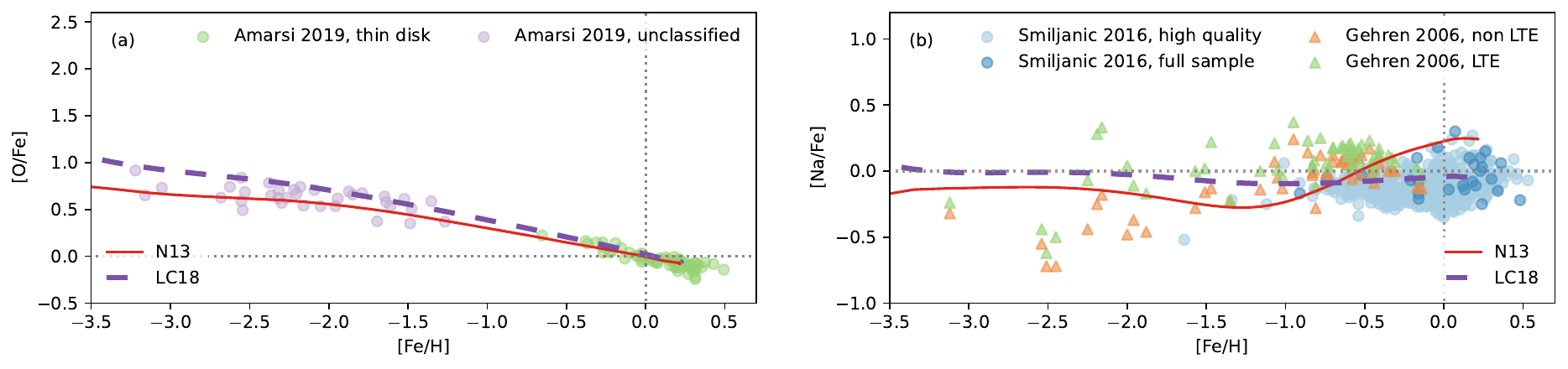} 
        \caption{Evolution tracks of $[\mathrm{O/Fe}]$ vs. $[\mathrm{Fe/H}]$
        (left) and $[\mathrm{Na/Fe}]$ vs. $[\mathrm{Fe/H}]$ (right) for the
        Solar neighbourhood. The red solid line shows the model adopting N13
        massive-star yields\citep[][]{nomoto2013}, while the purple dashed line
        adopts LC18 with \texttt{set~M} and \(v_{\rm
        rot}=0\,\mathrm{km\,s^{-1}}\); AGB yields are from \citet{karakas2010}.
        Grey dotted lines indicate the Solar abundances. Observational data are
        from \citet[][O]{amarsi2019}, \citet[][Na]{smiljanic2016}, and
        \citet[][Na]{gehren2006}. }
        \label{fig:MWmodel2}
\end{figure*}

Figure~\ref{fig:MWmodel2} compares the predicted $[\mathrm{O/Fe}]$ and
$[\mathrm{Na/Fe}]$ trends with observtions of stars in the Solar neighbourhood.
For oxygen we use the compilation of \citet{amarsi2019}, selecting only
thin-disc stars (green) and unclassified stars (purple). For sodium we adopt
data from \cite{smiljanic2016} and \cite{gehren2006}. In \citet{smiljanic2016},
data with a dispersion below 0.15~dex and derived from more than four
independent pipelines are classified as high-quality measurements (shown as
light-blue points), while the remaining data constitute the full sample (shown
as dark-blue points). In \citet{gehren2006}, abundances derived under the
assumption of local thermodynamic equilibrium (LTE) are labeled as LTE
measurements (orange points), whereas those obtained from non-LTE models are
classified as non-LTE measurements (green points). These data show a broadly
continuous decrease of $[\mathrm{O/Fe}]$ with increasing $[\mathrm{Fe/H}]$,
which is reasonably reproduced by our models.

Sodium measurements show substantial scatter at all metallicities. The model
adopting LC18 set~M (\(v_{\rm rot}=0\,\mathrm{km\,s^{-1}}\)) predicts
$[\mathrm{Na/Fe}]$ values that remain slightly sub-solar over
\([\mathrm{Fe/H}]\simeq-3.5\) to 0, in broad agreement with the average observed
trend. In contrast, the model adopting N13 yields shows an increase of
$[\mathrm{Na/Fe}]$ for \([\mathrm{Fe/H}]>-1\), which is not supported by the
bulk of the data. Given the uncertainties in Na nucleosynthesis and yield
uncertainties, both yield sets can be adopted for predicting chemical evolution
for star-forming galaxies. 

\section{Models of star-forming galaxies}
\label{sec:models of SF galaxies}

In this section, we extend our GCE model to star-forming galaxies with stellar
masses \(M_\star=10^{9}\text{--}10^{11}\,\mathrm{M}_\odot\). We select galaxies
that satisfy our selection criteria (described below) from GAEA semi-analytic
model, and use their star formation histories and present-day gas masses as
inputs to our calculations
\citep{delucia2014,delucia2024,hirschmann2016,xie2017,xie2020,fontanot2020}.
Given the good agreement reported in previous GAEA studies with a range of
observables \citep{delucia2020,Fontanot2021}, we adopt these model outputs as a
plausible representation of the SFHs of local star-forming galaxies.

The treatment of gas loss in GAEA and in NuPyCEE differs. We therefore treat the
mass-loading factor $\eta$ as a free parameter and calibrate it to
reproduce the observed \mz\ relation  \citep{Tremonti2004}.

Our fiducial model assumes the canonical Kroupa IMF \citep{kroupa2001} and N13
massive-star yields (Sect.~\ref{subsec:standard models}).  To study the effect
of IMF variations on sodium abundance in galaxies, we vary the IMF upper-mass
limit from 100 down to \(40\,\mathrm{M}_\odot\) (Sect.~\ref{subsec:diff IMF}).
For each IMF variant, we re-calibrate \(\eta\) such that the models remain
consistent with the observed \mz\, relation. Moreover, to assess the dependence
of our results on the adopted stellar yields, we repeat the analysis using the
LC18 \texttt{set~M} (\(v_{\rm rot}=0\,\mathrm{km\,s^{-1}}\)) yields. Additional
tests on the adopted gas masses, the SFH shape, and the IMF parameters are
presented in Appendices~\ref{app:Change gas mass}, \ref{app:Different SFH
type}, and \ref{app:Different alpha3}, respectively. A summary is given in
Table~\ref{tab: summary}.

\subsection{GAEA star formation histories}
\label{subsec:sfh}

The GAEA rendition considered in this work is based on the Millennium N-body
simulation \citep{Springel2005}. This model includes essential physical
processes of galactic evolution, such as gas accretion, cooling, star formation,
stellar feedback, AGN feedback, and galaxy mergers. This framework enables
tracing of the evolution of millions of galaxies, as well as their distribution
in the large-scale environment. GAEA is calibrated to reproduce many
observables, including the gas scaling relations at $z=0$, H{\sc i} and H$_2$
mass functions in the local Universe, the stellar mass function up to $z\sim 3$,
and the AGN luminosity function up to $z\sim 4$. Previous studies have
demonstrated that GAEA is able to reproduce the colour distribution and SFR
distribution of galaxies at $z=0$ \citep{delucia2024}, the local
gaseous-metallicity--galaxy-mass relation and its evolution \citep{delucia2020,
Fontanot2021}. We therefore use GAEA to prescribe SFHs and present-day gas
masses for our one-zone models. 

We select star-forming galaxies from GAEA in five stellar-mass bins
(\(\log(M_\star/\mathrm{M}_\odot)\pm0.1\)). Within each bin, we further require
the present-day SFR to be within \(\pm0.1\) dex of the star-forming main
sequence at \(z\simeq0\) as parameterised by \citet{Speagle2014}, and the
gas-phase oxygen abundance to fall within a broad range around the target value
(\(12+\log(\mathrm{O/H})\pm0.3\)). We then stack all the qualified GAEA galaxies
to obtain a smoothed, average SFH for each mass bin. We note that GAEA adopts a
Chabrier IMF \citep{chabrier2003}, with a star mass range of $[0.1,
100]\,\mathrm{M}_{\odot}$. Varying the upper limit of IMF affects the energy
release from stellar feedback and eventually impacts the SFH. Possible
implications of IMF-related inconsistencies between GAEA and our one-zone
post-processing are discussed in Sect.~\ref{sec:discussion}. 

\begin{figure}
    \centering
    \includegraphics[width=\columnwidth]{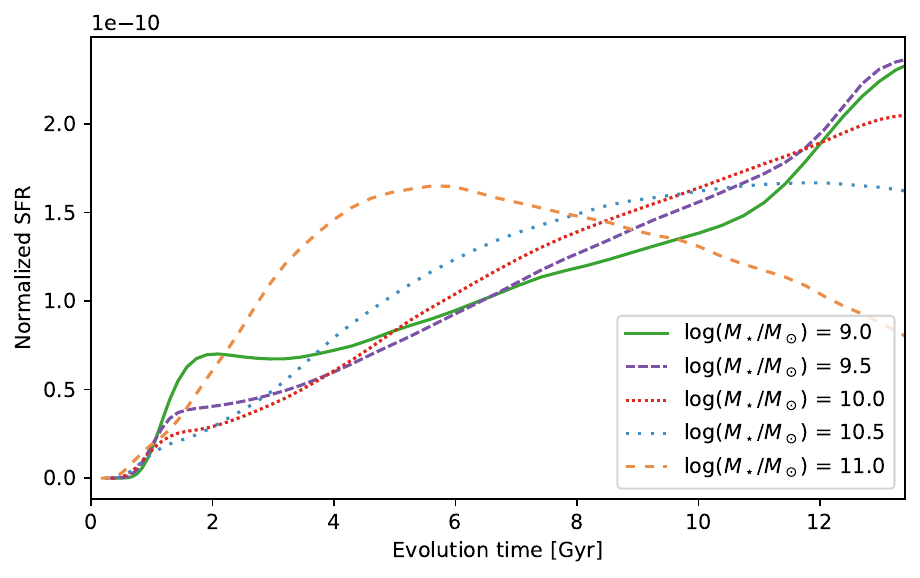}
    \caption{Average SFHs of star-forming galaxies in GAEA grouped in different
        mass bins and normalized to the respective stellar mass.}
    \label{fig: sfh}
\end{figure}

In total, over a thousand star-forming galaxies from GAEA are used to calculate
the average SFHs, shown in Fig.~\ref{fig: sfh}. The average SFHs are normalized
to the total stellar mass formed. Their shapes resemble the well-known
downsizing pattern, in which more massive galaxies form a larger fraction of
their stars at earlier times than that of lower-mass systems, in agreement with
trends inferred from observational data \citep{fontanot2009}.  Lower-mass
galaxies generally exhibit a rising SFR over time, whereas for the most massive
bins (e.g. \(\log(M_\star/\mathrm{M}_\odot)=10.5\) and 11), the SFR increases at
early times and then gradually declines.

\subsection{Standard models}
\label{subsec:standard models}

\begin{table}
    \caption{Final gas masses and $\eta$ parameter values for the
        standard models.}
    \label{tab: standard model}
    \centering
    \begin{threeparttable}
    \begin{tabular}{ccc}
    \hline \hline
         log(\mstar/\msun) & log(final gas mass/\msun)$^{a}$ & $\eta^{b}$  \\
    \hline
         9    & 9.1  & 1.5 \\
         9.5  & 9.5  & 0.9 \\
         10   & 9.8  & 0.4 \\
         10.5 & 10.1 & 0.3 \\
         11   & 10.3 & 0.6 \\
    \hline
    \end{tabular}
    \begin{tablenotes}
    \footnotesize
        \item $^{a}$ The final gas masses are from GAEA.
        \item $^{b}$ The $\eta$ parameter values are set to reproduce the
        observed \mz\ relation of \citet{Tremonti2004}.
    \end{tablenotes}
    \end{threeparttable}
\end{table}

\begin{figure}
    \centering
    \includegraphics[width=\columnwidth]{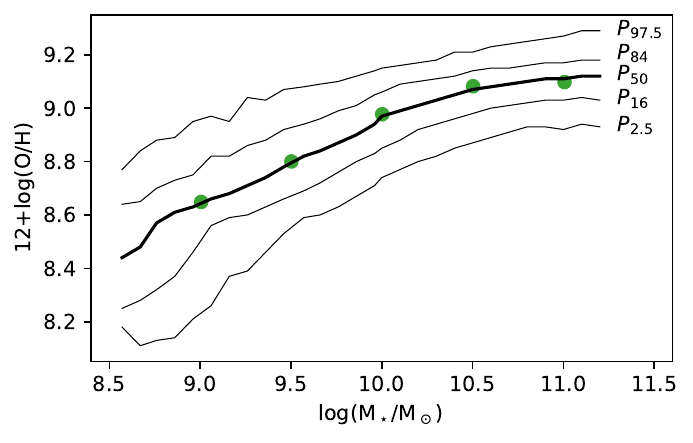}
    \caption{Final oxygen abundances of our standard models compared with the observed \mz\ relation from \citet{Tremonti2004}.}
    \label{fig: MZ relation}
\end{figure}

The adopted present-day gas masses (from GAEA) and the calibrated \(\eta\)
values for our standard models are listed in Table~\ref{tab: standard model}.
Given the present-day gas mass and SFR, the star-formation-law parameter
\(f_\star\) for each model is constrained. The final gas mass increases with
stellar mass, while the gas-to-stellar mass ratio decreases toward higher
stellar mass, in agreement with the observed trends \citep{Calura2008}. We
calibrate \(\eta\) to reproduce the observed gas-phase mass--metallicity
relation \citep{Tremonti2004}, as shown in Fig.~\ref{fig: MZ relation}. For
\(M_\star \lesssim 10^{11}\,\mathrm{M}_\odot\), the inferred \(\eta\) increases
toward lower masses, consistent with stronger effective gas loss in shallower
gravitational potentials \citep[e.g.][]{Pandya2021}. At the highest-mass bin,
the mass–metallicity relation approaches a plateau. In this regime,
the slower increase of oxygen abundance with stellar mass reduces the need for
$\eta$ to further decrease in order to maintain high metallicity. At the same
time, the elevated SFRs in massive galaxies can become a dominant factor in
driving gas loss, favoring larger $\eta$ values. These combined effects likely
explain the non-monotonic trend between $\eta$ and $M_\star$ in Table~\ref{tab:
standard model}.

\subsection{Models with different IMFs}
\label{subsec:diff IMF}

Varying the stellar IMF, while keeping other model parameters fixed,
significantly impacts the resulting galaxy properties. We vary the IMF
upper-mass limit \(m_{\rm up}\) between \(100\,\mathrm{M}_\odot\) and
\(40\,\mathrm{M}_\odot\). This variation has a negligible impact on the final
stellar mass because stars more massive than \(40\,\mathrm{M}_\odot\) have
lifetimes \(\lesssim 5\) Myr and account for \(\lesssim 6\%\) of the initial
stellar mass budget. Therefore, they evolve and return their mass to the
interstellar medium within a single time step, leaving the final stellar mass
largely unaffected.

In contrast, IMF variations can strongly affect the oxygen yield and thus the
final gas-phase oxygen abundance,  moving the models away from the observed
mass--metallicity relation. Since the observed \mz\ relation depends only weakly
on IMF assumptions, it represents a crucial external constraint for our models.
Therefore, we adjust the parameter \(\eta\) so that models with different
\(m_{\rm up}\) remain consistent with the \mz\ relation.

\begin{figure*}
    \centering
    \includegraphics[width=2\columnwidth]{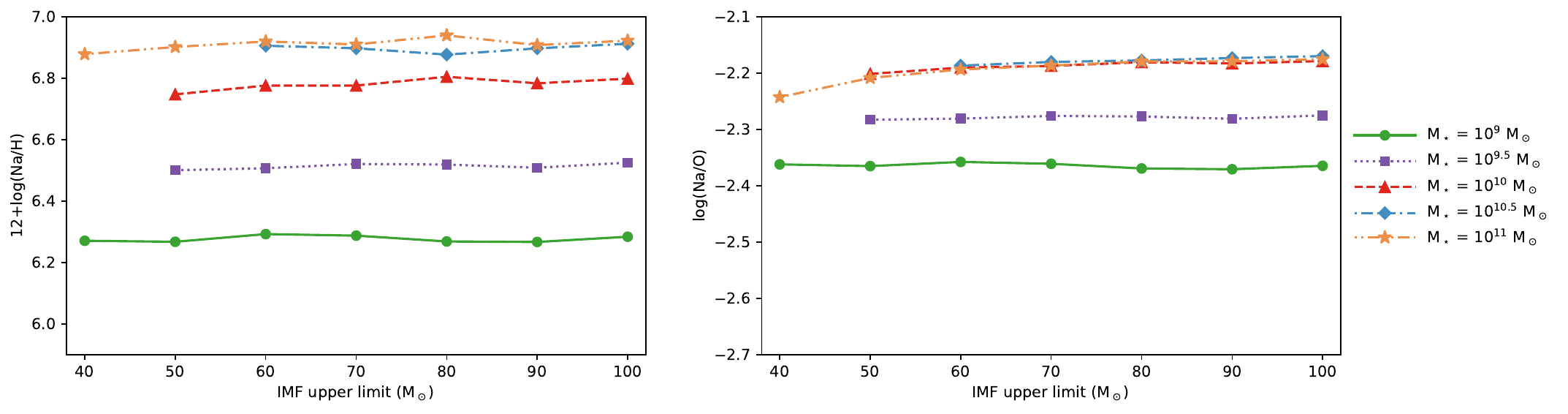}
    \caption{Final gas-phase sodium abundance and sodium-to-oxygen ratio of
        star-forming galaxy models with different IMF upper-mass limit \(m_{\rm
        up}\). Each model is re-calibrated to keep fitting the \mz\ relation.
        The adopted massive star yields are from N13. } \label{fig:
        n13-massloading}
\end{figure*}

\begin{figure*}
    \centering
    \includegraphics[width=2\columnwidth]{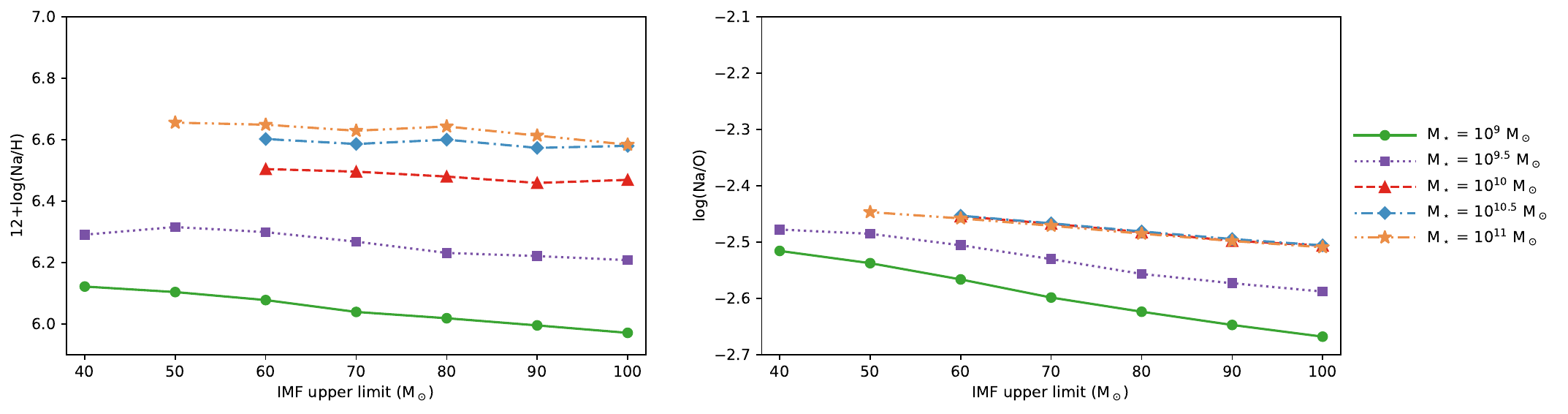}
    \caption{ Final gas-phase sodium abundance and sodium-to-oxygen ratio of
    star-forming galaxy models with different IMF upper-mass cutoffs $m_{\rm
    up}$. Each model is re-calibrated to keep fitting the \mz\ relation. The
    adopted massive star yields are from LC18, with \texttt{set~M} and $v =
    0\,\mathrm{km\,s^{-1}}$. Labels are the same as Fig \ref{fig:
    n13-massloading}.} 
    \label{fig: lc18-massloading}
\end{figure*}

Figure~\ref{fig: n13-massloading} shows the final gas-phase sodium abundance and
sodium-to-oxygen ratio of the modelled galaxies for different IMF upper-mass
limit \(m_{\rm up}\), after re-calibrating the mass-loading factor \(\eta\) to
reproduce the observed \mz\ relation (values listed in Table~\ref{tab:
Mup-massloading}). For both panels, for a given stellar mass, the spreads in
both 12+log(Na/H) and log(Na/O) are typically 0.1--0.2~dex, as a function of the
IMF upper limit. On the other hand, the variations between galaxy masses
(represented by different colours), for 12+log(Na/H) and log(Na/O), are $\sim
0.5$~dex  and $\sim 0.2$~dex, respectively. The stellar-mass dependence is
strongest at the low-mass end (\(\log(M_\star/\mathrm{M}_\odot)\lesssim 10\)),
while the high-mass bins show a smaller dynamic range. This behaviour indicates
that the final sodium abundance and the sodium-to-oxygen ratio are both
primarily determined by the galaxy's stellar mass through the imposed \mz\
relation, while the IMF upper mass limit plays an limited role. Since sodium
production is sensitive to metallicity, enforcing a tight constraint on O/H
largely limits the metallicity-dependent sodium production, leaving only a weak
residual dependence on \(m_{\rm up}\).

\subsection{Models with different stellar yields}
\label{subsec:diff stellar yield}

We repeat the same analysis using the LC18 \texttt{set~M} yields with \(v_{\rm
rot}=0\,\mathrm{km\,s^{-1}}\). The re-calibrated \(\eta\) values needed to
reproduce the \mz\ relation, as well as the final sodium abundances, are listed
in Table~\ref{tab: Mup-lc18-massloading}. The resulting trends of sodium
abundance and sodium-to-oxygen ratio are shown in Fig.~\ref{fig:
lc18-massloading}. Compared to the N13-based models, we find a more sensitive
response to \(m_{\rm up}\) in the lowest-mass bin
(\(\log(M_\star/\mathrm{M}_\odot)=9.0\)). This is consistent with the
IMF-weighted yields in Fig.~\ref{fig: yield}, which indicate that at low
metallicity a larger fraction of sodium can originate from the most massive
stars (\(\gtrsim 40\,\mathrm{M}_\odot\)) in the LC18 \texttt{set~M} yields.

However, even with this heightened dependence, the overall variation in the
final sodium abundance across the tested \(m_{\rm up}\) range remains small
\(\lesssim 0.2\)~dex. The \(\log(\mathrm{Na/O})\) (right panel of Fig.~\ref{fig:
lc18-massloading}) shows a clearer trend with \(m_{\rm up}\), but its full
dynamic range is still confined to \(\lesssim 0.2\)~dex. Given typical
systematic uncertainties in gas-phase abundance measurements, such variations
are unlikely to provide a strong observational constraint on \(m_{\rm up}\).

\section{Discussion and conclusion}
\label{sec:discussion}

As noted in Section~\ref{subsec:sfh}, our analysis combines SFHs extracted from
the GAEA semi-analytical model with GCE model that adopts a different IMF. In
GAEA, a Chabrier IMF is assumed over the mass range $[0.1,
100]\,\mathrm{M}_\odot$, whereas our GCE calculations employ a Kroupa-type IMF
with a default range of $[0.01, 100]\,\mathrm{M}_\odot$. More than that, we also
explore the effects of hypernova explosions on nucleosynthesis, which are not
included in GAEA energy budgets and, consequently, the resulting SFHs. To
quantify this effect, we compare the number of massive stars ($m >
8\,\mathrm{M}_\odot$) formed per unit stellar mass, the Kroupa IMF adopted here
causes a reduction in the massive stars around $11\%$ compared with the Chabrier
IMF used in GAEA. Even in an extreme case where the upper mass limit of Kroupa
IMF is reduced to $40\,\mathrm{M}_\odot$, the corresponding reduction remains
around $14\%$. These differences only introduce minor variations in the
integrated stellar feedback energy budget, and are therefore not expected to
significantly alter the SFHs adopted in our modelling. In addition, the SFHs
used in this work are normalised to the final stellar mass and only their shapes
are adopted; as such, differences in the IMF do not affect the final stellar
mass. We therefore conclude that the impact of the IMF mismatch on our results
is limited.

In our modelling, the mass-loading factor is treated as a free parameter to
reproduce the observed gas-phase metallicity and stellar mass at the final
evolutionary stage. For models sharing the same stellar mass and SFH but
different IMF upper limits, the resulting mass-loading factors are different.
Changing only the IMF upper limit from 100 to $40\,\rm{M}\odot$ reduces the
number of massive stars by approximately $3.5\%$, which is unlikely to
significantly affect the energy budget from SNe and stellar winds. Moreover, the
mass-loading factor has a complex connection with galaxy properties.
Theoretically, properties such as SNe clustering, SFR surface density, and gas
surface density can influence the mass-loading factor\citep{Kim2020, Smith2021,
Deng2024}, while these properties are not included in our zero-dimensional
model. Observationally, the mass-loading factor also suffers from large
uncertainties \citep[e.g.,][]{McQuinn2019, Concas2022, XrismCollaboration2026}.
In this sense, our adopted values serve as a reference, pending more robust
constraints from future measurements, which will help refine the model design.

\begin{figure*}
    \centering
    \includegraphics[width=2\columnwidth]{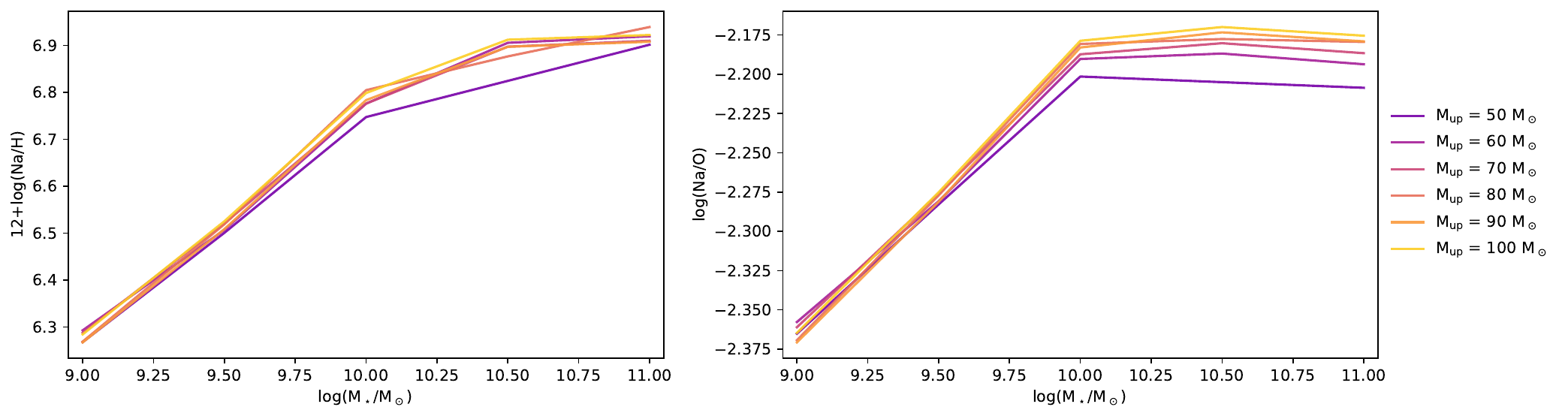}
    \caption{Final sodium abundance and final sodium-to-oxygen abundance ratio
        of our model galaxies as a function of \mstar, for different choices of
        $m_{\rm{up}}$. The adopted massive star yields are from N13.}
        \label{fig: na_mstar}
\end{figure*}

\begin{figure*}
    \centering
    \includegraphics[width=2\columnwidth]{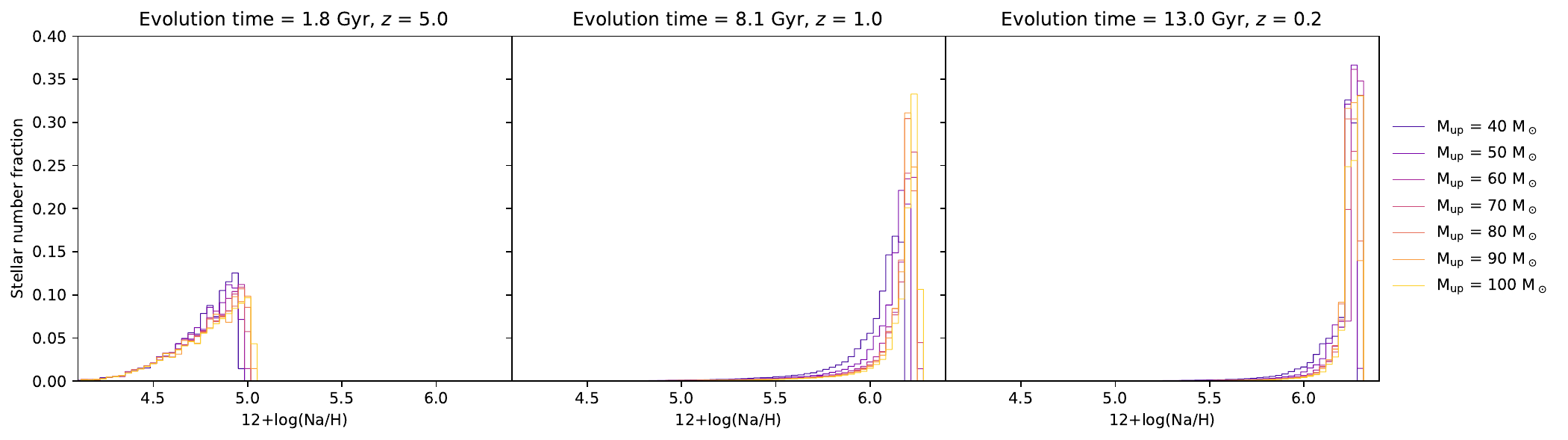}
    \caption{Number fraction distribution of stellar sodium abundances at
    different evolutionary epochs (corresponding redshifts are given in
    Fig.~\ref{fig: sfh}) for models with log(\mstar/\msun) = 9.0 and different
    \Mup. The adopted massive star yields are from N13.} 
    \label{fig: na_number_distribution}
\end{figure*}

As shown in the previous section, the sodium abundance in star-forming galaxies
is only weakly sensitive to changes in the IMF upper-mass limit \(m_{\rm up}\),
but it correlates strongly with galaxy stellar mass, largely driven by the \mz\
relation. The resulting correlations between sodium abundance (and
sodium-to-oxygen ratio) and \(M_\star\) are shown in Fig.~\ref{fig: na_mstar}.
Both 12+log(Na/H) and log(Na/O) increase as galaxy mass increases from
\(M_\star\sim 10^{9}\,\mathrm{M}_\odot\) to \(10^{11}\,\mathrm{M}_\odot\), and
the trend flattens around \(M_\star\sim 10^{10}\,\mathrm{M}_\odot\), consistent
with the corresponding flattening in the \mz\ relation.

From these models, sodium production is dominated by massive stars, while LIMS
contribute at most \(\sim 10\%\) of the total sodium budget (see Fig.~\ref{fig:
yield}). Because the same massive-star populations also dominate oxygen
production, enforcing the \mz\ relation largely sets O/H and therefore the
metallicity-dependent sodium production. As a consequence, the predicted
log(Na/H)--\(M_\star\) trend closely follows the \mz\ relation, and varying
\(m_{\rm up}\) introduces a negligible role. 

In stellar population synthesis, sodium absorption features are widely involved
under assumptions of the IMF and SFH for unresolved stellar populations
\citep{labarbera2013}. However, when these assumptions are changed, the derived
physical properties, e.g.,  stellar mass, abundances for both stellar and
gas-phases, are also affected. To construct the integrated spectrum of a galaxy,
it is necessary to combine contributions from different stellar populations.
Determining the distribution of stellar abundances formed at different times is
critical for accurate population synthesis studies. 

In Fig.~\ref{fig: na_number_distribution}, we show the fraction of stars with
different sodium abundances at various redshifts, which would be the input of a
GCE-based self-consistent spectra synthesis models. The sodium abundance of newly
formed stars increases gradually with cosmic time. Therefore, within our
modelling framework and the explored range of \(m_{\rm up}\) (indicated by
different colours), the overall shape of the distribution changes slightly, but
no significant systematic differences or clear trends are observed. Therefore,
the change of IMF upper limit is likely to have only a minor effect on the
sodium-sensitive integrated features in stellar population synthesis.

In this work, our GCE predictions for sodium are based on the widely adopted
stellar yield compilations from \citet{karakas2010} for AGB stars and
\citet{nomoto2013} or \citet{limongi2018} for massive stars. While these models
provide a robust baseline for interpreting the relative insensitivity of
$^{23}$Na to variations in the high-mass IMF, we must acknowledge that the
absolute abundance scales are subject to systematic uncertainties stemming from
the underlying nuclear physics inputs and yields of stellar evolution models.  
On the nuclear physics side, a primary source of uncertainty for $^{23}$Na
production in AGB stars is the $^{22}$Ne($p,\gamma$)$^{23}$Na reaction rate,
which governs sodium synthesis during the Hot Bottom Burning phase. As
demonstrated by \citet{Williams2020}, this rate has historically carried
non-negligible uncertainties. Their inverse kinematics study highlights that
while experimental precision has improved, the reaction rate discrepancies
between different databases (e.g.,
NACRE\footnote{https://www.astro.ulb.ac.be/nacreii/index.html} vs.
STARLIB\footnote{https://starlib.github.io/Rate-Library/}) and different
experiment groups (e.g., TUNL\footnote{https://nucldata.tunl.duke.edu/},
DRAGON\footnote{https://dragon.triumf.ca/home.html}, and
LUNA\footnote{https://luna.lngs.infn.it/}) can still span orders of magnitude at
low temperature.

For massive stars, the uncertainties in the $^{12}$C+$^{12}$C fusion reaction
rate significantly impact the production of $^{23}$Na during carbon burning.
\citet{Tang2022} emphasises that the total $^{12}$C+$^{12}$C reaction
cross-section in stellar interior remains constrained by the competition between
potential resonance structures and hindrance effects.  Furthermore, relative
production of $^{23}$Na is sensitive to the branching ratio of $^{12}$C+$^{12}$C
fusion between the p-exit ($^{23}$Na$+p$) and $\alpha$-exit ($^{20}$Ne$+\alpha$)
channels.  \citet{DeGeronimo2024} highlights that contrary to the constant value
assumed historically, this branching ratio is temperature-dependent and can
deviate significantly. Because this ratio directly modulates how much
carbon-fusion flow feeds the $^{23}$Na production, it can systematically shift
the final $^{23}$Na yields. However,  as noted in the review by
\citet{Chieffi2025}, the experimental difficulty of disentangling carbon-fusion
products at low energies like the stellar interiors remains remarkable. Future
direct experiments at low energies in underground facilities as LUNA
\citep[Laboratory for Underground Nuclear Astrophysics,][]{LUNA} and
JUNA\citet[Jingpin Underground Nuclear Astrophysics collaboration][]{JUNA} may
help to further constrain $^{23}$Na production in different stars from the
nuclear physics perspective.  

A major source of uncertainty for AGB yields, particularly for sodium, is the
treatment of the third dredge-up (TDU). The efficiency and onset of this
convective mixing process, which brings newly synthesised material (including
sodium) to the surface, are highly model-dependent. Variations in its
parameterisation can alter predicted sodium yields by an order of magnitude or
more, especially at low metallicities where the neutron-to-seed ratio is high
\citep{karakas2010}. For massive stars, our adopted yields do not account for
stellar rotation. Rotation-induced mixing can significantly enhance the
production of the neutron source $^{22}$Ne in the helium-burning shell, thereby
increasing sodium yields, particularly in the low-metallicity environments where
rotational velocities are hypothesised to be higher \citep{limongi2018}.
Consequently, our models, which use non-rotating massive star yields, may
systematically underestimate early sodium enrichment.

Future work incorporating improved nuclear data and a self-consistent set of
yields that include rotation and a detailed exploration of convective boundary
mixing will be crucial to refine the absolute timeline and scaling of galactic
sodium enrichment.

\begin{acknowledgement}
We thank Prof. Alessandro Bressan for useful discussions. We acknowledge the
support of the National Natural Science Foundation of China (NSFC) under grants
No. 1257030642, 12533003, 12203021, 12573040, 12533008, and 12203100. We
acknowledge the Postdoctoral Fellowship Program of CPSF under Grant Number
GZC20252097. TJ acknowledges the support from the MUNI Award in Science and
Humanities(MUNI/I/1762/2023). We acknowledge the support of Tiangong
International Overseas Exchange Scholarship. 

\end{acknowledgement}

\bibliographystyle{aa}
\bibliography{main.bib}

\begin{appendix}

\section{Summary for all the model sets in this work}

\newcounter{num}

\begin{table*}
    \caption{The summary of all the parameters and inputs we have used in this work.} 
    \label{tab: summary}
    \centering
    \begin{threeparttable}
    \begin{tabular}{llllllll}
        \hline \hline 
        Model set & Section & yield & IMF variation$^{a}$ & SFH & $\eta$ & \mgas & Result \\
        \hline
        \setcounter{num}{1}\Roman{num} & Section~\ref{subsec:diff IMF}        & N13  & \Mup       & star-forming & free$^{b}$     & GAEA & Fig.~\ref{fig: n13-massloading} \\
        \setcounter{num}{2}\Roman{num} & Section~\ref{subsec:diff IMF}        & LC18 & \Mup       & star-forming & free     & GAEA & Fig.~\ref{fig: lc18-massloading} \\
        \setcounter{num}{3}\Roman{num} & Section~\ref{app:Change gas mass}    & N13  & \Mup       & star-forming & standard & free & Fig.~\ref{fig: n13-gas} \\
        \setcounter{num}{4}\Roman{num} & Section~\ref{app:Different SFH type} & N13  & \Mup       & quenching    & free     & GAEA & Fig.~\ref{fig: quench-ml} \\
        \setcounter{num}{5}\Roman{num} & Section~\ref{app:Different alpha3}   & N13  & $\alpha_3$ & star-forming & free     & GAEA & Fig.~\ref{fig: alpha3-n13} \\
        \hline 
    \end{tabular}
    \begin{tablenotes}
    \small
    \item $^{a}$ This means which part of IMF we changed to test weather the IMF will influence the final sodium abundance. 
    \item $^{b}$ This means this parameter is set to free parameter to match with \mz\ relation. 
    \end{tablenotes}
    \end{threeparttable}
\end{table*}

Table~\ref{tab: summary} provides an overview of all model sets explored in this
work, together with their key input parameters and corresponding results. For
each model set, we list the section where it is discussed, the adopted stellar
yield, the type of IMF variation, the assumed star formation history (SFH), the
treatment of the mass-loading factor $\eta$, and the gas mass prescription. The
last column indicates the figures in which the main results of each model set
are presented. This table is intended to serve as a roadmap for the paper,
allowing readers to quickly identify the assumptions and outcomes of each
experiment.

\section{Fit the \mz\ relation by changing mass-loading factor}
\begin{table}
    \caption{Mass loading factors ($\eta$) and final sodium abundance in the models$^{a}$.}
    \label{tab: Mup-massloading}
    \centering
    \begin{threeparttable}
    \begin{tabular}{cccc}
    \hline \hline
        log(\mstar/\msun) & \Mup(\msun) & $\eta^{b}$ & 12+log(Na/H)$^{c}$ \\
    \hline
        9.0   & 40   & 0.3  & 6.3  \\ 
        9.0   & 50   & 0.8  & 6.3  \\ 
        9.0   & 60   & 1.0  & 6.3  \\ 
        9.0   & 70   & 1.2  & 6.3  \\ 
        9.0   & 80   & 1.4  & 6.3  \\ 
        9.0   & 90   & 1.5  & 6.3  \\ 
        9.0   & 100  & 1.5  & 6.3  \\ 
        9.5   & 40   & - $^*$    & -    \\ 
        9.5   & 50   & 0.4  & 6.5  \\ 
        9.5   & 60   & 0.6  & 6.5  \\ 
        9.5   & 70   & 0.7  & 6.5  \\ 
        9.5   & 80   & 0.8  & 6.5  \\ 
        9.5   & 90   & 0.9  & 6.5  \\ 
        9.5   & 100  & 0.9  & 6.5  \\ 
        10.0  & 40   & -    & -    \\ 
        10.0  & 50   & 0.1  & 6.7  \\ 
        10.0  & 60   & 0.2  & 6.8  \\ 
        10.0  & 70   & 0.3  & 6.8  \\ 
        10.0  & 80   & 0.3  & 6.8  \\ 
        10.0  & 90   & 0.4  & 6.8  \\ 
        10.0  & 100  & 0.4  & 6.8  \\ 
        10.5  & 40   & -    & -    \\ 
        10.5  & 50   & -    & -    \\ 
        10.5  & 60   & 0.1  & 6.9  \\ 
        10.5  & 70   & 0.2  & 6.9  \\ 
        10.5  & 80   & 0.3  & 6.9  \\ 
        10.5  & 90   & 0.3  & 6.9  \\ 
        10.5  & 100  & 0.3  & 6.9  \\ 
        11.0  & 40   & 0.1  & 6.9  \\ 
        11.0  & 50   & 0.3  & 6.9  \\ 
        11.0  & 60   & 0.4  & 6.9  \\ 
        11.0  & 70   & 0.5  & 6.9  \\ 
        11.0  & 80   & 0.5  & 6.9  \\ 
        11.0  & 90   & 0.6  & 6.9  \\ 
        11.0  & 100  & 0.6  & 6.9  \\ 
    \hline
    \end{tabular}
    \begin{tablenotes}
    \small
    \item $^{a}$ In this set of models, we adopt the N13 yield for massive stars and star-forming type SFH.
    \item $^{b}$ $\eta$ in this table is adopted to reproduce the \mz\ relation for galaxies with different stellar masses and different IMF
    upper-mass limits ($m_{\rm up}$).
    \item $^{c}$ $12+\log({\rm Na/H})$ represents the final sodium abundance for every model.
    \item $^*$ Rows marked with ``-'' indicate models that could not be calibrated to the observed \mz\ relation.  
    \end{tablenotes}
    \end{threeparttable}
\end{table}

In Table~\ref{tab: Mup-massloading}, we show the $\eta$ we adopt for every model
to match the \mz\ relation. For the same stellar mass and decreasing upper limit
of IMF, the model has to adopt a lower $\eta$ to reduce the gas loss to lock more
oxygen in the ISM. For the final sodium abundance, the galaxies with the same
stellar mass and different IMF \Mup\ seems have quite similar final sodium
abundance, while the galaxies with the same IMF \Mup\ and increasing stellar
mass tend to have higher final sodium abundance. This is detailed described and
discussed in Section~\ref{subsec:diff IMF}.

\section{Fit the \mz\ relation with LC18 yield}
\label{app:Different stellar yield}

\begin{table}
    \caption{Mass loading factors ($\eta$) and final sodium abundance in the models$^{a}$.}
    \label{tab: Mup-lc18-massloading}
    \centering
    \begin{threeparttable}
    \begin{tabular}{cccc}
    \hline \hline
        log(\mstar/\msun) & \Mup(\msun) & $\eta^{b}$ & 12+log(Na/H)$^{c}$ \\
    \hline
        9.0   & 40   & 0.3  & 6.1  \\ 
        9.0   & 50   & 0.7  & 6.1  \\ 
        9.0   & 60   & 1.0  & 6.1  \\ 
        9.0   & 70   & 1.3  & 6.0  \\ 
        9.0   & 80   & 1.5  & 6.0  \\ 
        9.0   & 90   & 1.7  & 6.0  \\ 
        9.0   & 100  & 1.9  & 6.0  \\ 
        9.5   & 40   & 0.1  & 6.3  \\ 
        9.5   & 50   & 0.3  & 6.3  \\ 
        9.5   & 60   & 0.5  & 6.3  \\ 
        9.5   & 70   & 0.7  & 6.3  \\ 
        9.5   & 80   & 0.9  & 6.2  \\ 
        9.5   & 90   & 1.0  & 6.2  \\ 
        9.5   & 100  & 1.1  & 6.2  \\ 
        10.0  & 40   & - $^*$   & -    \\ 
        10.0  & 50   & -    & -    \\ 
        10.0  & 60   & 0.2  & 6.5  \\ 
        10.0  & 70   & 0.3  & 6.5  \\ 
        10.0  & 80   & 0.4  & 6.5  \\ 
        10.0  & 90   & 0.5  & 6.5  \\ 
        10.0  & 100  & 0.5  & 6.5  \\ 
        10.5  & 40   & -    & -    \\ 
        10.5  & 50   & -    & -    \\ 
        10.5  & 60   & 0.1  & 6.6  \\ 
        10.5  & 70   & 0.2  & 6.6  \\ 
        10.5  & 80   & 0.2  & 6.6  \\ 
        10.5  & 90   & 0.3  & 6.6  \\ 
        10.5  & 100  & 0.3  & 6.6  \\ 
        11.0  & 40   & -    & -    \\ 
        11.0  & 50   & 0.2  & 6.7  \\ 
        11.0  & 60   & 0.3  & 6.6  \\ 
        11.0  & 70   & 0.4  & 6.6  \\ 
        11.0  & 80   & 0.4  & 6.6  \\ 
        11.0  & 90   & 0.5  & 6.6  \\ 
        11.0  & 100  & 0.6  & 6.6  \\ 
    \hline
    \end{tabular}
    \begin{tablenotes}
    \small
    \item $^{a}$ In this set of models, we adopt the LC18 yield for massive stars and star-forming type SFH.
    \item $^{b}$ $\eta$ adopted to reproduce the \mz\ relation for galaxies with different stellar masses and different IMF upper-mass limits ($m_{\rm up}$).
    \item $^{c}$ 12+log(Na/H) represents the final sodium abundance for every model.
    \item $^*$ Rows marked with ``-'' indicate models that could not be calibrated to the observed \mz\ relation.  
    \end{tablenotes}
    \end{threeparttable}
\end{table}

In Table~\ref{tab: Mup-lc18-massloading}, we show the same results as in
Table~\ref{tab: Mup-massloading}. Although the specific value change slighly,
the total trend of the $\eta$ and final sodium abundance in this table is quite
similar within the Table~\ref{tab: Mup-lc18-massloading}, so we believe that the
stellar yield do not influence our conclusion a lot. The trend of the final
sodium abundance is also shown in Fig.~\ref{fig: lc18-massloading}. Detailed
analysis of final sodium abundance evolution is shown in Section~\ref{subsec:diff IMF}.

\section{Fit \mz\ relation by changing gas mass}
\label{app:Change gas mass}

When we change the IMF \Mup\ in a model, the oxygen abundance of this model will
be decreased and deviate from the \mz\ relation. In the main text, we use
smaller $\eta$ to reduce the gas loss and match the model to \mz\ relation.
However, this might introduce bias to our results, so we also test to use other
method to match the \mz\ relation. In this Section, after change the IMF \Mup,
we change the final gas mass in the model to fit with the \mz\ relation.

\begin{table}[]
    \caption{Final gas mass (\mgas) and final sodium abundance in the models$^{a}$.}
    \label{tab: Mup-gas}
    \centering
    \begin{threeparttable}
    \begin{tabular}{cccc}
    \hline \hline
        log(\mstar/\msun) & \Mup(\msun) & log(\mgas/\msun)$^{b}$ & 12+log(Na/H)${c}$ \\
    \hline
        9.0   & 40   & - $^*$   & -    \\ 
        9.0   & 50   & -    & -    \\ 
        9.0   & 60   & 8.6  & 6.3  \\ 
        9.0   & 70   & 8.9  & 6.3  \\ 
        9.0   & 80   & 9.0  & 6.3  \\ 
        9.0   & 90   & 9.1  & 6.3  \\ 
        9.0   & 100  & 9.1  & 6.3  \\ 
        9.5   & 40   & -    & -    \\ 
        9.5   & 50   & -    & -    \\ 
        9.5   & 60   & 8.5  & 6.5  \\ 
        9.5   & 70   & 9.2  & 6.5  \\ 
        9.5   & 80   & 9.4  & 6.5  \\ 
        9.5   & 90   & 9.4  & 6.5  \\ 
        9.5   & 100  & 9.5  & 6.5  \\ 
        10.0  & 40   & -    & -    \\ 
        10.0  & 50   & -    & -    \\ 
        10.0  & 60   & 7.8  & 6.8  \\ 
        10.0  & 70   & 9.4  & 6.8  \\ 
        10.0  & 80   & 9.6  & 6.8  \\ 
        10.0  & 90   & 9.7  & 6.8  \\ 
        10.0  & 100  & 9.8  & 6.8  \\ 
        10.5  & 40   & -    & -    \\ 
        10.5  & 50   & -    & -    \\ 
        10.5  & 60   & -    & -    \\ 
        10.5  & 70   & 9.4  & 6.9  \\ 
        10.5  & 80   & 9.8  & 6.9  \\ 
        10.5  & 90   & 10.1 & 6.9  \\ 
        10.5  & 100  & 10.1 & 6.9  \\ 
        11.0  & 40   & -    & -    \\ 
        11.0  & 50   & -    & -    \\ 
        11.0  & 60   & -    & -    \\ 
        11.0  & 70   & -    & -    \\ 
        11.0  & 80   & -    & -    \\ 
        11.0  & 90   & 10.3 & 6.9  \\ 
        11.0  & 100  & 10.3 & 6.9  \\ 
    \hline
    \end{tabular}
    \begin{tablenotes}
    \small
    \item $^{a}$ In this set of models, we adopt the N13 yield for massive stars and star-forming type SFH.
    \item $^{b}$ \mgas\ adopted to reproduce the \mz\ relation for galaxies with different stellar masses and different IMF upper-mass limits ($m_{\rm up}$).
    \item $^{c}$ 12 + log(Na/H) represents the final sodium abundance of every model.
    \item $^*$ Rows marked with ``-'' indicate models that could not be calibrated to the observed \mz\ relation.  
    \end{tablenotes}
    \end{threeparttable}
\end{table}

\begin{figure*}
    \centering
    \includegraphics[width=2\columnwidth]{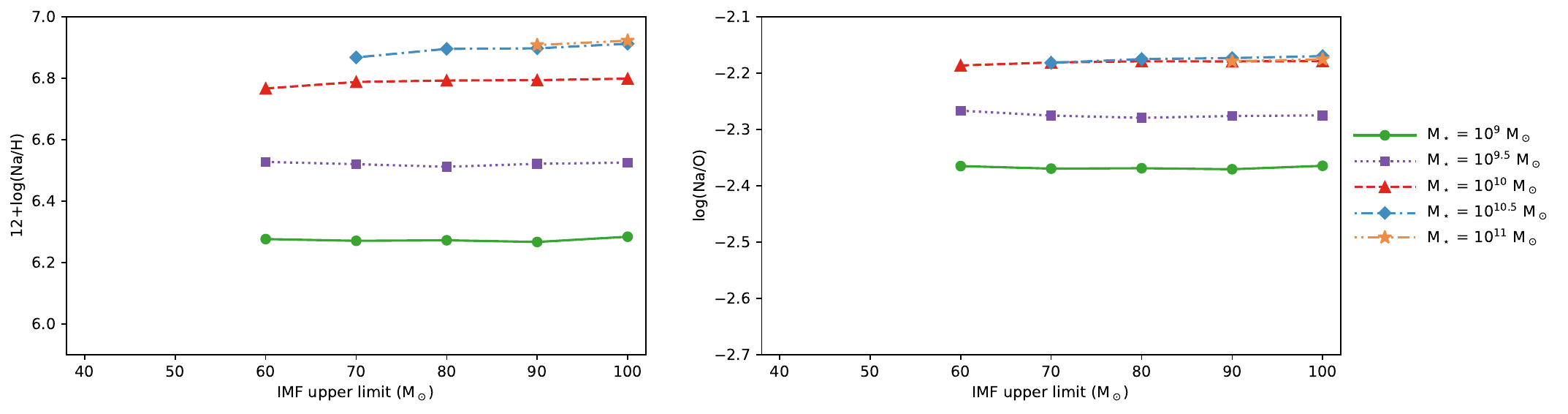}
    \caption{Final sodium abundance and final sodium-to-oxygen ratio of model
    star-forming galaxies with different IMF's upper mass limits. The models are
    calibrated so as to keep fitting the \mz\ relation. The adopted massive star
    yields are from N13. }
    \label{fig: n13-gas}
\end{figure*}

In Table~\ref{tab: Mup-gas}, we show the final gas mass we adopt for every model
to match the \mz\ relation. For the same stellar mass and decreasing upper limit
of IMF, the model has to adopt a lower gas mass to enhance the oxygen abundance
in the galaxy. For the final sodium abundance, the galaxies with the same
stellar mass and different IMF \Mup\ seems have quite similar final sodium
abundance, while the galaxies with the same IMF \Mup\ and increasing stellar
mass tend to have higher final sodium abundance. 

Fig.~\ref{fig: n13-gas} shows the final sodium abundance (left panel) and the
final sodium-to-oxygen ratio (right panel) as a function of the IMF \Mup, for
galaxies with different stellar masses. Overall, both 12+log(Na/H) and log(Na/O)
exhibit weak dependence on \Mup\ for all stellar mass bins, which is consistent
with the results in the Fig.~\ref{fig: n13-massloading}. Thus, we believe that
the different method of matching the \mz\ relation will not influence our final
conclusion.

\section{Different SFH type}
\label{app:Different SFH type}

\begin{figure}
    \centering
    \includegraphics[width=\columnwidth]{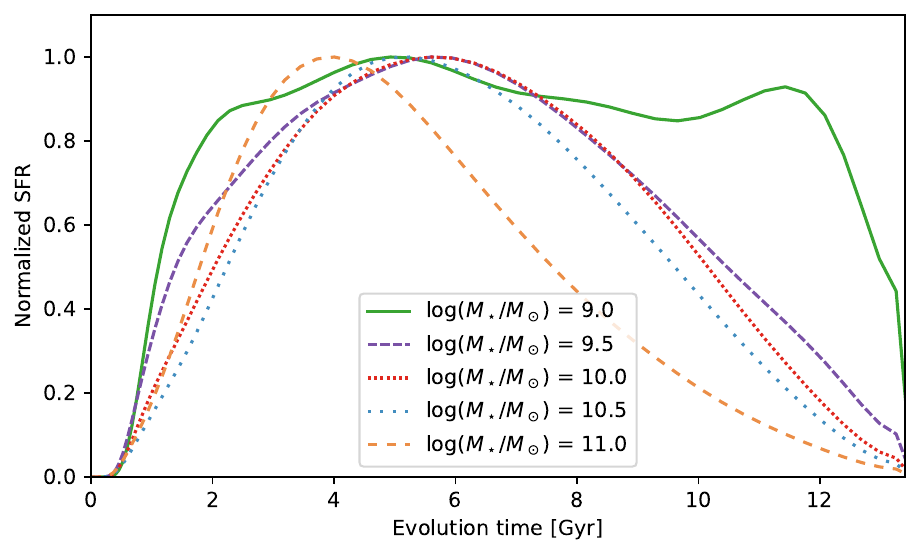}
    \caption{Average SFHs of quenching galaxies in GAEA grouped in different mass bins and normalized to the respective stellar mass (see text).}
    \label{fig: quench_sfh}
\end{figure}

To avoid the SFH bias to our results, we also use a quenching type of SFH to do
the same comparison. As in Section~\ref{subsec:sfh}, we select over a thousand
quenched galaxies from GAEA and compute their smoothed average SFHs. Despite the
final SFR has to reach 0, other criteria are still kept to the same. The final
SFH we get are shown in Fig.~\ref{fig: quench_sfh}, and they are all normalised
to their stellar masses.

\begin{table}[]
    \centering
    \caption{Mass loading factors ($\eta$) and final sodium abundance in the models$^{a}$.}
    \label{tab: Mup-quench}
    \begin{threeparttable}
    \begin{tabular}{cccc}
    \hline \hline
        log(\mstar/\msun) & \Mup(\msun) & $\eta^{b}$ & 12+log(Na/H)$^{c}$ \\
    \hline 
        9.0   & 40   & 3.7  & 6.1  \\ 
        9.0   & 50   & 3.9  & 6.2  \\ 
        9.0   & 60   & 3.9  & 6.2  \\ 
        9.0   & 70   & 3.9  & 6.2  \\ 
        9.0   & 80   & 3.9  & 6.2  \\ 
        9.0   & 90   & 3.9  & 6.3  \\ 
        9.0   & 100  & 3.9  & 6.3  \\ 
        9.5   & 40   & 1.6  & 6.2  \\ 
        9.5   & 50   & 2.0  & 6.2  \\ 
        9.5   & 60   & 2.3  & 6.2  \\ 
        9.5   & 70   & 2.5  & 6.2  \\ 
        9.5   & 80   & 2.6  & 6.2  \\ 
        9.5   & 90   & 2.7  & 6.2  \\ 
        9.5   & 100  & 2.8  & 6.2  \\ 
        10.0  & 40   & 0.6  & 6.3  \\ 
        10.0  & 50   & 0.8  & 6.3  \\ 
        10.0  & 60   & 1.0  & 6.3  \\ 
        10.0  & 70   & 1.1  & 6.3  \\ 
        10.0  & 80   & 1.1  & 6.3  \\ 
        10.0  & 90   & 1.2  & 6.3  \\ 
        10.0  & 100  & 1.2  & 6.3  \\ 
        10.5  & 40   & 0.2  & 6.5  \\ 
        10.5  & 50   & 0.4  & 6.5  \\ 
        10.5  & 60   & 0.5  & 6.5  \\ 
        10.5  & 70   & 0.6  & 6.5  \\ 
        10.5  & 80   & 0.7  & 6.5  \\ 
        10.5  & 90   & 0.7  & 6.5  \\ 
        10.5  & 100  & 0.7  & 6.5  \\ 
        11.0  & 40   & 0.2  & 6.5  \\ 
        11.0  & 50   & 0.3  & 6.6  \\ 
        11.0  & 60   & 0.4  & 6.5  \\ 
        11.0  & 70   & 0.5  & 6.5  \\ 
        11.0  & 80   & 0.5  & 6.5  \\ 
        11.0  & 90   & 0.6  & 6.5  \\ 
        11.0  & 100  & 0.6  & 6.5  \\ 
    \hline
    \end{tabular}
    \begin{tablenotes}
    \small
        
    \item $^{a}$ In this set of models, we adopt the N13 yield for massive stars and quenching type SFH.
    \item $^{b}$ $\eta$ adopted to reproduce the \mz\ relation for galaxies with different stellar masses and different IMF upper-mass limits ($m_{\rm up}$).
    \item $^{c}$ 12 + log(Na/H) represents the final sodium abundance for every model.
    \end{tablenotes}
    \end{threeparttable}
\end{table}

\begin{figure*}
    \centering
    \includegraphics[width=2\columnwidth]{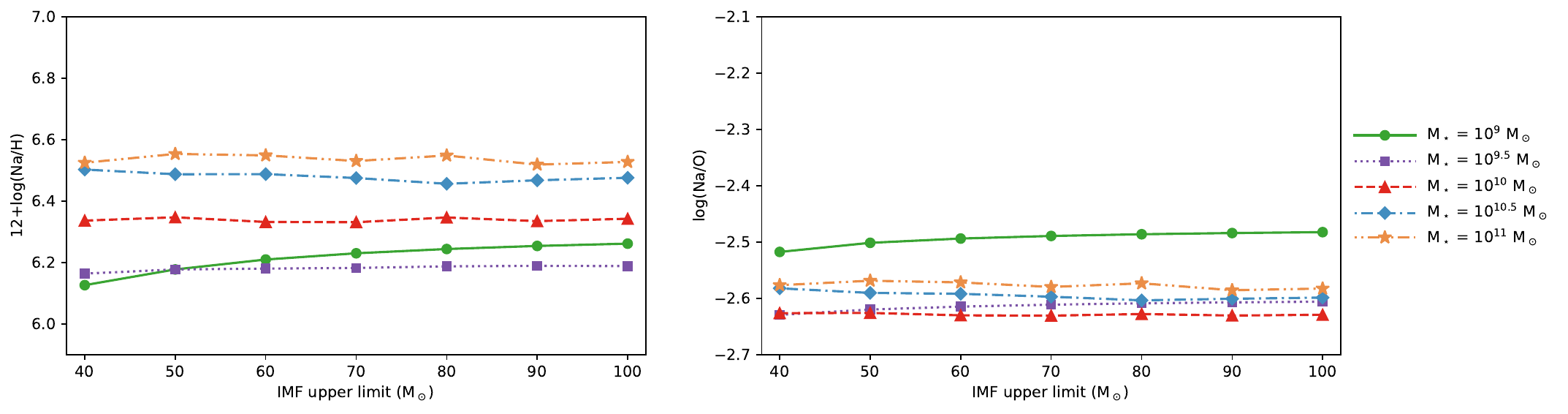}
    \caption{Final sodium abundance and final sodium-to-oxygen ratio of model
    quenching galaxies with different IMF's upper mass limits. The models are
    calibrated so as to keep fitting the \mz\ relation. The adopted massive star
    yields are from N13. }
    \label{fig: quench-ml}
\end{figure*}

Table~\ref{tab: Mup-quench} lists the mass-loading factors, $\eta$, adopted for
each model to reproduce \mz\ relation. Overall, the trends of $\eta$ and the
final sodium abundance are quite similar to those shown in Table~\ref{tab:
Mup-massloading}. However, for models with the same stellar mass and the same
IMF upper mass limit, \Mup, different SFHs require different $\eta$, especially
at the low-mass end. E.g., for galaxies with $M_\star=10^9\,\mathrm{M}_\odot$
and \Mup=$100\,\mathrm{M}_\odot$, the $\eta$ adopted for the quenching SFH is
more than a factor of two higher than that for the star-forming SFH. This
difference likely arises because, although the final surviving stellar masses
are the same, quenching galaxies form more stars in total than star-forming
galaxies. As a result, they produce more oxygen and therefore require a larger
mass-loading factor to enhance gas loss in order to match the oxygen abundance
implied by the \mz\ relation.

Figure~\ref{fig: quench-ml} shows the final sodium abundance (left panel) and
the final sodium-to-oxygen ratio (right panel) as a function of the IMF \Mup,
for galaxies with different stellar masses. Overall, both 12+log(Na/H) and
log(Na/O) exhibit weak dependence on \Mup\ for all stellar mass bins. For a
given stellar mass, the variation in sodium abundance is within ~0.1 dex when
\Mup\ changes from 40 to $100\,\mathrm{M}_\odot$, consistent with our previous
conclusion. In contrast, a clear stellar-mass dependence is observed. 

\section{Different $\alpha_3$}
\label{app:Different alpha3}

In this section, we use different IMF slope to test the influence of IMF to
sodium abundance. We set the $\alpha_3$ of the IMF to be -2.5, -2.4, -2.3, -2.2,
-2.1, and -2.0. As in standard model setting, we change the $\eta$ of the model
to fit with the \mz\ relation. 

\begin{table}[]
    \caption{Mass loading factors ($\eta$) and final sodium abundance in the models$^{a}$.}
    \label{tab: alpha3-n13}
    \centering
    \begin{threeparttable}
    \begin{tabular}{cccc}
    \hline \hline
        log(\mstar/\msun) & $\alpha_3$ & $\eta^{b}$ & 12+log(Na/H)$^{c}$ \\
    \hline
        9.0   & -2.0 & 4.9  & 6.3  \\ 
        9.0   & -2.1 & 3.8  & 6.3  \\ 
        9.0   & -2.2 & 2.6  & 6.3  \\ 
        9.0   & -2.3 & 1.5  & 6.3  \\ 
        9.0   & -2.4 & 0.7  & 6.3  \\ 
        9.0   & -2.5 & 0.1  & 6.3  \\ 
        9.5   & -2.0 & 3.5  & 6.5  \\ 
        9.5   & -2.1 & 2.5  & 6.5  \\ 
        9.5   & -2.2 & 1.7  & 6.5  \\ 
        9.5   & -2.3 & 0.9  & 6.5  \\ 
        9.5   & -2.4 & 0.3  & 6.5  \\ 
        9.5   & -2.5 & - $^*$   & -    \\ 
        10.0  & -2.0 & 2.1  & 6.8  \\ 
        10.0  & -2.1 & 1.5  & 6.8  \\ 
        10.0  & -2.2 & 0.9  & 6.8  \\ 
        10.0  & -2.3 & 0.4  & 6.8  \\ 
        10.0  & -2.4 & -    & -    \\ 
        10.0  & -2.5 & -    & -    \\ 
        10.5  & -2.0 & 1.7  & 6.9  \\ 
        10.5  & -2.1 & 1.2  & 6.9  \\ 
        10.5  & -2.2 & 0.7  & 6.9  \\ 
        10.5  & -2.3 & 0.3  & 6.9  \\ 
        10.5  & -2.4 & -    & -    \\ 
        10.5  & -2.5 & -    & -    \\ 
        11.0  & -2.0 & 1.7  & 7.0  \\ 
        11.0  & -2.1 & 1.3  & 7.0  \\ 
        11.0  & -2.2 & 0.9  & 7.0  \\ 
        11.0  & -2.3 & 0.6  & 6.9  \\ 
        11.0  & -2.4 & 0.3  & 6.9  \\ 
        11.0  & -2.5 & -    & -    \\ 
    \hline
    \end{tabular}
    \begin{tablenotes}
    \small
    \item $^{a}$ In this set of models, we adopt the N13 yield for massive stars and star-forming type SFH.
    \item $^{b}$ $\eta$) adopted to reproduce the \mz\ relation for galaxies with different stellar masses and different IMF slope.
    \item $^{c}$ 12 + log(Na/H) represents the final sodium abundance for every model.
    \item $^*$ Rows marked with ``-'' indicate models that could not be calibrated to the observed \mz\ relation.  
    \end{tablenotes}
    \end{threeparttable}
\end{table}

\begin{figure*}
    \centering
    \includegraphics[width=2\columnwidth]{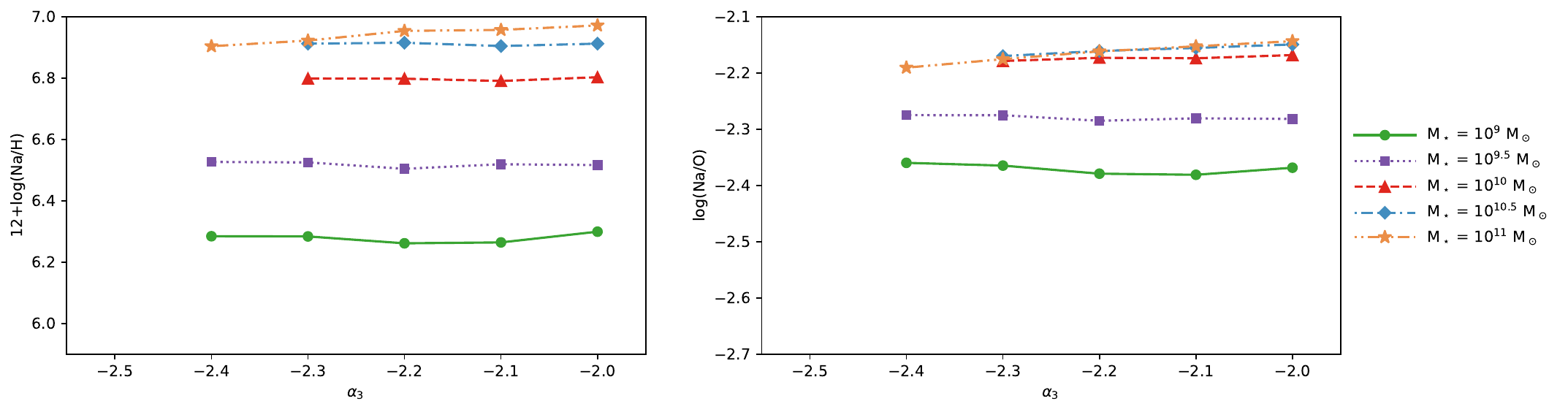}
    \caption{Final sodium abundance and final sodium-to-oxygen ratio of model
    star-forming galaxies with different IMF slope. The models are calibrated so
    as to keep fitting the \mz\ relation. The adopted massive star yields are
    from N13. }
    \label{fig: alpha3-n13}
\end{figure*}

The $\eta$ we use and the final sodium abundance we get are listed in
Table~\ref{tab: alpha3-n13}, and the final results are shown in Figure~\ref{fig:
alpha3-n13}. For all models, the $\alpha_3$ could not be -2.6, or any slope
which is steeper than -2.5. Otherwise, the galaxies could not match with \mz\ 
relation. The pattern of final sodium abundance is quite similar with the
pattern in Figure~\ref{fig: n13-massloading}, with the IMF becoming more
top-heavy, the final sodium abundance slightly change. This is the same for the
final sodium-to-oxygen ratio. Therefore, no matter we change the IMF upper limit
or change the slope of the massive part of IMF, the final sodium abundance are
not sensitive to the change of IMF.

\section{Loose the constraint of oxygen abundance}

\begin{figure*}
    \centering
    \includegraphics[width=2\columnwidth]{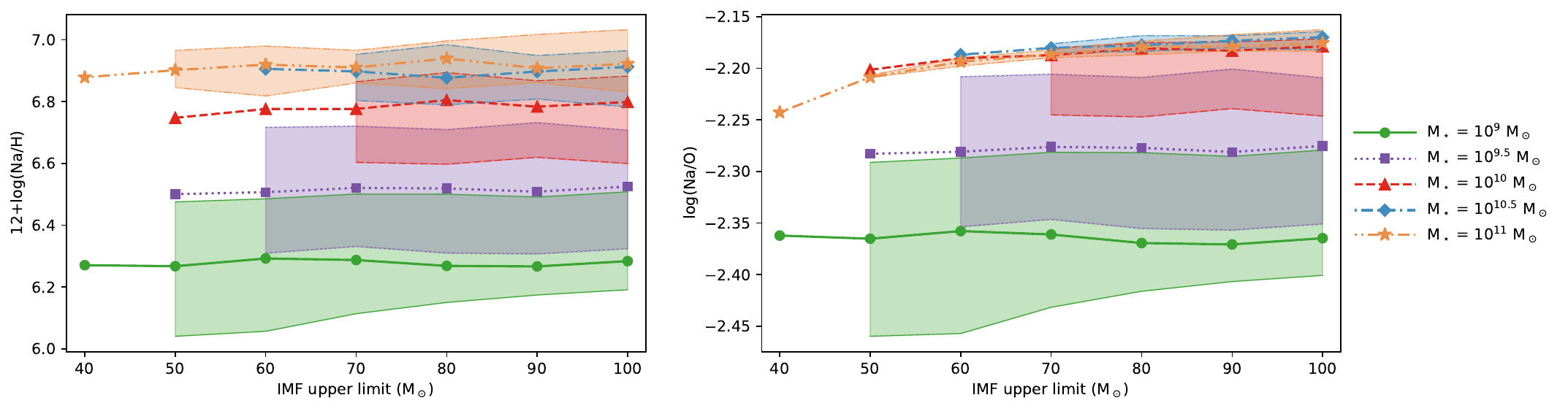}
    \caption{Final sodium abundance and final sodium-to-oxygen ratio of model
    star-forming galaxies with different IMF \Mup. The models are calibrated so
    as to keep fitting the \mz\ relation. The adopted massive star yields are
    from N13. }
    \label{fig: n13-ml-region}
\end{figure*}

In our fiducial models, we strictly impose the oxygen abundance constraint from
the \mz\ relation. However, not all galaxies in Universe can perfectly match
this relation. We therefore test whether relaxing the oxygen abundance
constraint affects our conclusions. Specifically, we extend the allowed oxygen
abundance range from the median ($P_{50}$) to the $P_{16}$--$P_{84}$ interval,
as shown in Fig.~\ref{fig: MZ relation}. The corresponding ranges of the final
sodium abundance (left panel) and sodium-to-oxygen ratio (right panel) are shown
in Fig.~\ref{fig: n13-ml-region}. As shown in this figure, when the oxygen
abundance constraint is relaxed by $\sim$0.2 dex, both 12+log(Na/H) and
log(Na/O) vary by a similar amount. We find that this level of variation does
not affect our main conclusions.

\end{appendix}
\end{document}